\documentclass[a4paper,fleqn,usenatbib,useAMS]{mnras}

\usepackage{amssymb}	
\usepackage{multicol}        
\usepackage{bm}		
\usepackage{pdflscape}	
\usepackage[T1]{fontenc}
\usepackage{ae,aecompl}

\usepackage {graphicx}
\usepackage{natbib}
\usepackage{aas_journals}
\usepackage{color}
\usepackage{amsmath}
\definecolor{mygray}{gray}{0.7}

\def\aap{{ A\&A}}

\def\apjl{{ApJL}}

\def\apjs{{ApJS}}

\newcommand{\bes}{\begin{equation*}}
\newcommand{\ees}{\end{equation*}}
\newcommand{\bea}{\begin{eqnarray}}
\newcommand{\eea}{\end{eqnarray}}
\newcommand{\beas}{\begin{eqnarray*}}
\newcommand{\eeas}{\end{eqnarray*}}

\newcommand{\ltsima}{$\; \buildrel < \over \sim \;$}
\newcommand{\lsim}{\lower.5ex\hbox{\ltsima}}
\newcommand{\gtsima}{$\; \buildrel > \over \sim \;$}
\newcommand{\gsim}{\lower.5ex\hbox{\gtsima}}

\def\gtrsim{\mathrel{\hbox{\rlap{\hbox{\lower4pt\hbox{$\sim$}}}\hbox{$>$}}}}
\let\ga=\gtrsim
\def\lesssim{\mathrel{\hbox{\rlap{\hbox{\lower4pt\hbox{$\sim$}}}\hbox{$<$}}}}
\let\la=\lesssim

\definecolor{mygray}{gray}{0.5}

\newcommand{\be}{\begin{equation}}
\newcommand{\ee}{\end{equation}}
\newcommand{\ba}{\begin{eqnarray}}
\newcommand{\ea}{\end{eqnarray}}

\title[Cold imprint of voids and \emph{super}voids]{The part and the whole: voids, \emph{super}voids, and their ISW imprint}
\author[Andr\'as Kov\'acs]{Andr\'as Kov\'acs\thanks{akovacs@ifae.es}\\
Institut de F\'isica d'Altes Energies, The Barcelona Institute of Science and Technology, E-08193 Bellaterra (Barcelona), Spain}

\begin{document}
\date{Submitted 2017}
\pagerange{\pageref{firstpage}--\pageref{lastpage}} \pubyear{2017}
\maketitle
\label{firstpage}
\begin{abstract}
The integrated Sachs-Wolfe imprint of extreme structures in the cosmic web probes the dynamical nature of dark energy. Looking through typical cosmic voids, no anomalous signal has been reported. On the contrary, \emph{super}voids, associated with large-scale fluctuations in the gravitational potential, have shown potentially disturbing excess signals. In this study, we used the Jubilee ISW simulation to demonstrate how the stacked signal depends on the void definition. We found that large underdensities, with at least $\approx5$ merged sub-voids, show a peculiar ISW imprint shape with central cold spots and surrounding hot rings, offering a natural way to define \emph{super}voids in the cosmic web. We then inspected the real-world BOSS DR12 data using the simulated imprints as templates. The imprinted profile of BOSS \emph{super}voids appears to be more compact than in simulations, requiring an extra $\alpha \approx 0.7$ re-scaling of filter sizes. The data reveals an excess ISW-\emph{like} signal with $A_{\rm ISW}\approx9$ amplitude at the $\approx2.5\sigma$ significance level, unlike previous studies that used isolated voids and reported good consistency with $A_{\rm ISW}=1$. The tension with the Jubilee-based $\Lambda$CDM predictions is $\gsim2\sigma$, in consistency with independent analyses of \emph{super}voids in Dark Energy Survey data. We show that such a very large enhancement of the $A_{\rm ISW}$ parameter hints at a possible causal relation between the CMB Cold Spot and the Eridanus \emph{super}void. The origin of these findings remains unclear.

\end{abstract}
\section{Introduction}
A dynamical property of dark energy is the decay of large-scale gravitational potentials which imprint tiny secondary anisotropies to the primary fluctuations of the Cosmic Microwave Background (CMB) radiation. This late-time re-procession of CMB patterns by the cosmic web is studied in the framework of the Integrated Sachs-Wolfe effect \citep[ISW]{SachsWolfe} in the linear regime, and via the subdominant Rees-Sciama effect \citep[RS]{ReesSciama} on smaller scales. The weak ISW imprints on the primordial CMB temperature fluctuations can be measured in cross-correlations with tracers of the matter distribution \citep{CrittendenTurok1996}. 

Summarizing the individual measurement efforts, \cite{gian} (see also references therein) combined several tracer catalogues and reported an $A_{\rm ISW}=\Delta T_{\rm data} / \Delta T_{\rm \Lambda CDM}\approx1.38\pm0.32$ ``amplitude" using angular cross-correlation techniques, where $A_{\rm ISW}=1$ corresponds to the concordance $\Lambda$-Cold Dark Matter ($\Lambda$CDM) prediction. With more tracer catalogues, the {\it Planck} team also combined the individual attempts and found $A_{\rm ISW}\approx1.00\pm0.25$ in cross-correlation functions \citep{PlanckISW2015}.

Alternatively, the ISW signal may also be detected \emph{locally} using catalogues of voids and superclusters. The measurement involves the identification of individual voids in the cosmic web assuming a void definition and then a stacking of CMB temperatures on their locations as a measure of their average imprint. Typically, no high-significance detection has been reported \citep{Ilic2013,Planck19,CaiEtAl2014,Hotchkiss2015,Kovacs2015} with differently constructed void catalogues in the Sloan Digital Sky Survey (SDSS) data using the \texttt{ZOBOV} algorithm \citep{ZOBOV}. These studies all allowed some level of void merging using \texttt{ZOBOV}'s watershed method, but specific analyses of potentially encompassing \emph{super}voids have not been attempted directly.

Recently, \cite{NadathurCrittenden2016} reported a $3.1\sigma$ detection of the ISW signal from ``isolated" voids and superclusters in the Baryon Oscillations Spectroscopic Survey (BOSS) data release 12 (DR12). Their implementation of the watershed algorithm prevented neighboring voids from merging \citep[see][]{NadathurEtal2016}. They used optimal matched filters and found $A_{\rm ISW}\approx1.65\pm0.53$. 

Based on the same BOSS tracer data set but a different void catalogue, \cite{Cai2016} found $A_{\rm ISW}\approx6$ with a marginal $\approx1.6\sigma$ detection significance. However, using those voids seen to be the most probable with $p_{\rm void}>3\sigma$ (i.e., least likely to occur in random catalogues), they found $A_{\rm ISW}\approx20$ at $\approx3.4\sigma$ significance. The imprint of these voids showed no anomaly in the {\it Planck} CMB lensing convergence ($\kappa$) map. 

Both studies focussed on efficient pruning strategies to, above all, remove the so-called voids-in-clouds that are expected to be aligned  with hot spots on the CMB. Apart from the different filtering methods applied, most importantly \cite{Cai2016} also considered merged voids. In part, this difference might explain the different outcomes because \cite{Hotchkiss2015} have pointed out in simulations that the shape of the stacked ISW imprint does depend on the void definition.

Based on stacking probes with systems of merged sub-voids or \emph{super}voids, however, there is another branch of observational results that reported $A_{\rm ISW}\approx10$ values \citep[using SDSS, Pan-STARRS1\footnote{http://pswww.ifa.hawaii.edu/pswww/}, and Dark Energy Survey (DES) data, respectively]{GranettEtAl2008,SzapudiEtAl2014,Kovacs2016}. In this paper, we test these claims using yet another type of BOSS DR12 void catalogue. We focus on systems of merged voids using so called ``minimal" voids \citep{Nadathur2015} that trace large-scale underdensities in the cosmic web. See Table 1 for examples of void selection criteria. 

We use the Jubilee ISW simulation \citep{Watson2014} and a mock luminous red galaxy (LRG) catalogue to {\it a priori} define pruning strategies. We aim to understand the differences between ISW measurement techniques that consider isolated voids and \emph{super}voids. Finally, we study the implications of our findings to the problem of the CMB Cold Spot \citep{CruzEtal2004} and the Eridanus \emph{super}void \citep{SzapudiEtAl2014}.

The paper is organized as follows. In Section 2, we discuss the role of void definitions in the currently available observational ISW landscape. Data sets and detection algorithms are introduced in Section 3. Our simulated and observational results are presented in Section 4 and in Section 5, respectively, while the final section contains a summary, discussion and interpretation of our findings.

\begin{table}
\centering
\caption{Strategies for controlling merging in the void hierarchy. See Nadathur \& Hotchkiss (2015) for details.}
\begin{tabular}{@{}ccc}
\hline
\hline
Label &  \multicolumn{2}{c}{Criteria for merging:} \\
 & Link density & Density ratio \\
\hline
VIDE & $n_\rmn{link}<0.2\overline{n}$ & unconstrained \\
Minimal & $n_\rmn{link}<\overline{n}$ & $r<2$ \\
Isolated & no merging & no merging \\
\hline
\hline
\end{tabular}
\label{table:merging}
\end{table}

\section{ISW in photo-$z$ catalogues}

In photometric data, finding typical voids surrounded by overdensitites is challenging because of the smearing effect of photo-$z$ errors in the line-of-sight (LOS) distribution of galaxies. Systems of voids lined up in our LOS, however, are possible to detect with new algorithms and a good understanding of void properties and potential biases in void identification \citep{Sanchez2016}.

If accounted for, this selection effect might actually be an advantage because super-structures elongated in our LOS have a longer photon travel time compared to the spherical case, corresponding to larger ISW temperature shifts. However, it is worth noting that \cite{Flender2013} concluded that the assumption of sphericity does not lead to a significant underestimate of the ISW signal in a $\Lambda$CDM model.

Nevertheless, the higher-than-expected ISW-like signals seem to emerge when using catalogues of this kind of merged voids based on tracers affected by photo-$z$ errors.

\begin{figure*}
\begin{center}
\includegraphics[width=160mm]{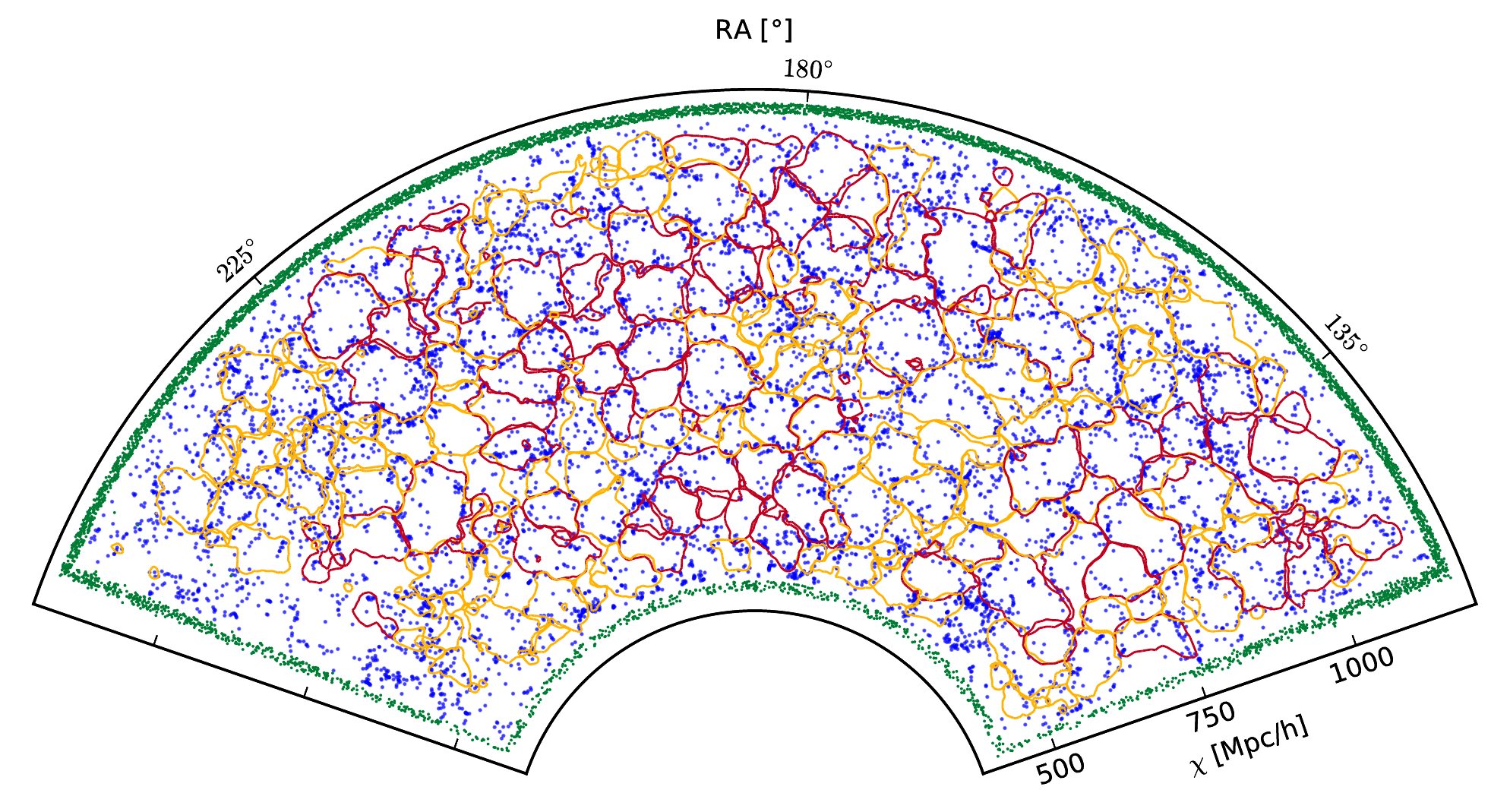}
\caption{The cross-section of voids in the LOWZ sample of the BOSS DR12 data with the cone $\rmn{Dec}=12^\circ$, coloured according to whether the average galaxy density within the void is $\overline\delta_g<0$ (red) or $\overline\delta_g>0$ (yellow). Galaxy positions in a slice of opening angle $2^\circ$ centred at the angle are overlaid in blue, and buffer mocks around the survey edges in green. Voids with $\overline\delta_g<0$ tend to correspond to under-compensated underdensities, forming \emph{super}void features, while those with $\overline\delta_g>0$ are on average over-compensated on large scales, i.e. voids-in-clouds. Minimal voids can be defined by merging these isolated voids under specific criteria. (Plot from Nadathur (2016).)}
\end{center}
\label{para}
\end{figure*}

\subsection{SDSS \emph{super}voids and their ISW\emph{-like} effect}
Foremost, \cite{GranettEtAl2008} (Gr08, hereafter) defined a catalog of 50-50 significant \emph{super}voids and superclusters using SDSS photo-$z$ data. Using a BOSS DR12 spec-$z$ galaxy catalogue, \cite{Granett2015} reconstructed the average redshift space shape of the Gr08 \emph{super}voids that were originally defined by photo-$z$ tracers, finding an axis ratio $R_{\parallel}/R_{\perp} \approx 2.6\pm 0.4$.

For this sample, Gr08 found a higher-than-expected ISW\emph{-like} signal that appears to be in $\approx3\sigma$ tension with $\Lambda$CDM predictions with $A_{\rm ISW}\approx10$ \citep{PapaiSzapudi2010,PapaiEtAl2011,Nadathur2012,Flender2013,Aiola}. The freedom to vary the $\Lambda$CDM parameters, given other constraints, is not enough to overcome the discrepancy with observation.

Besides, \cite{Hernandez2013} found that varying the number of the objects in the stacking lowers the overall significance. The simulation analyses by \cite{Kovacs2016} have shown, however, that stacking all voids in a catalogue might not be the optimal strategy for the highest signal-to-noise ($S/N$, hereafter) detection of the ISW imprint. The largest voids have bigger impact but they are less numerous therefore an optimum might exist halfway; perhaps close to the serendipitous choice by Gr08. 

Another problem with the original Gr08 is the {\it a posteriori} choice of filter size for their compensated top-hat filter (hereafter CTH, see section 4.1 for details). Gr08 used a constant $R=4^{\circ}$ filter size but the re-scaling of filters to the individual void size appears to be important. For the Gr08 \emph{super}voids, a filter re-scaling of $R/R_{v}\approx0.6$ for $\tilde{R}_{v}$ void radii maximized the signal \citep{Ilic2013}. This is in line with the simulation analyses by \cite{CaiEtAl2014}.

In summary, the original Gr08 signal has survived new CMB data releases and tests against CMB and galactic systematics and remains a puzzle. It is important to look for similar signals elsewhere in the sky to test the hypothesis of a rare statistical fluctuation.

\subsection{DES \emph{super}voids and their ISW\emph{-like} effect}

Recently, \cite{Kovacs2016} probed the Gr08 claims with photo-$z$ data in a different footprint. They used the first year data of the Dark Energy Survey \citep[DES,][]{DES}. They identified 52 large voids and 102 superclusters at redshifts $0.2 < z < 0.65$ using the void finder tool described in \cite{Sanchez2016}. The heart of that method is a restriction to 2D slices of galaxy data, and measurements of the projected density field around centers defined by minima in the corresponding smoothed density field. A larger smoothing automatically merges smaller sub-voids into larger voids, while too coarse smoothing can increase the uncertainties in the position and size estimates.

They then tested the shapes and orientations of their super-structures. Analyses of DES mock galaxy catalogues revealed a mean LOS elongation $R_{\parallel}/R_{\perp} \approx 2.2$ for voids and $R_{\parallel}/R_{\perp} \approx 2.6$ for superclusters in redshift space. In contrast, for voids in the BOSS spec-$z$ data \cite{Nadathur2016} found smaller average ellipticities, and with a random orientation of void major axes relative to the LOS.

In their analysis, \cite{Kovacs2016} used the Jubilee simulation to {\it a priori} evaluate the configuration. Following Gr08 and others, they performed a stacking measurement with the CTH filtering technique. For optimal configurations, they found a cold imprint of voids with $A_{\rm ISW}\approx8\pm6$ that is $\approx1.2\sigma$ higher than the imprint of such super-structures in Jubilee's $\Lambda$CDM universe. They also found $A_{\rm ISW}\approx8\pm5$ for their superclusters. If they instead used an {\it a posteriori} selected filter size $R/R_{v}=0.6$, they found $A_{\rm ISW}\approx15$ which exceeds $\Lambda$CDM expectations at the $\approx2\sigma$ level. Note that Gr08 and DES \emph{super}void catalogues both show elongation along the line-of-sight and for both samples the $R/R_{v}\approx0.6$ re-scaling maximizes their ISW-like imprint with a similarly high $A_{\rm ISW}$ amplitude.

\section{Data sets for the ISW analysis}
\begin{figure*}
\begin{center}
\includegraphics[width=172mm]{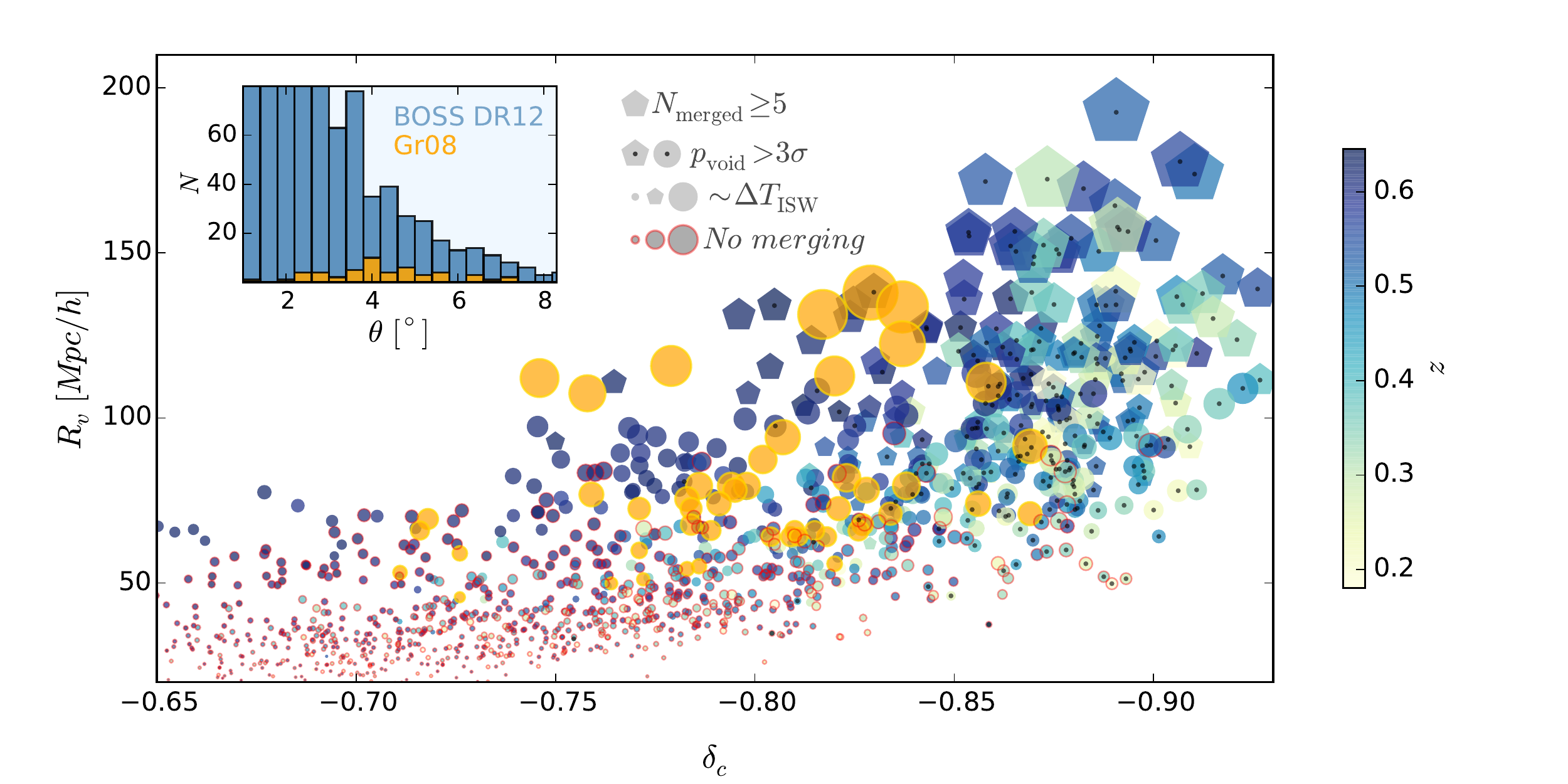}
\caption{Summary plot of minimal void parameters $\tilde{R}_{v}$ radius, $\delta_{c}$ central density, and $z$ redshift in the BOSS DR12 CMASS and LOWZ joint catalogue. Circle sizes indicate relative central ISW expectations. Orange circles mark Gr08 \emph{super}voids. The inset shows the angular size distributions. Pentagon symbols show \emph{super}voids with at least 5 merged sub-voids (used later for final conclusions), while red-edged circles mark objects without merging. Black points in the centers of circles or pentagons indicate $p_{\rm void}>3\sigma$ void probability.}
\end{center}
\label{para}
\end{figure*}

\subsection{The Eridanus \emph{super}void: ISW\emph{-like} effect?}

The CMB Cold Spot \citep{CruzEtal2004} is one of the large-scale anomalies in the CMB. Its significance is $\approx2-3\sigma$ depending on the statistical method applied. \cite{Nadathur2014} argued that it is not its coldness that makes it anomalous but the combination of a rather cold area in the centre and a surrounding hot ring feature.

Nevertheless there is growing observational evidence, again in photo-$z$ data, for the presence of the low-$z$ Eridanus \emph{super}void almost perfectly aligned with the Cold Spot \citep{RudnickEtal2007,SmithHuterer2010,GranettEtal2010,BremerEtal2010,SzapudiEtAl2014}. Similarly to SDSS and DES \emph{super}voids, the Eridanus \emph{super}void was found to be significantly elongated in the LOS \citep{KovacsJGB2015} and it is a rare matter fluctuation. Its exact shape is not yet known but it appears to be a complicated system of sub-voids  \citep{Mackenzie2017}.

Assuming viable void profiles, analytical models predict ISW imprint profiles for this \emph{super}void that disagree with the observed profile of the Cold Spot \citep{Nadathur2014,FinelliEtal2014,MarcosCaballero2015,KovacsJGB2015,Naidoo2016,Naidoo2017}. Relatedly, \cite{NadathurCrittenden2016} concluded that the ISW explanation is not supported by their results because a very large enhancement of the $A_{\rm ISW}$ parameter would be required.

The significance of these observational findings is unclear at the moment. Nevertheless a trend might be emerging in the shadow of {\it a posteriori} bias arguments, indicating that the globally estimated $A_{\rm ISW}\approx1$ amplitude might have a larger value for the largest super-structures. We test this hypothesis with special samples of merged voids in BOSS DR12 data and in the Jubilee simulation.

\subsection{Isolated voids}
The concept of isolated voids does not include merging of voids into \emph{super}voids (see Figure 1). The problem with merging, as \cite{Nadathur2016} discussed, is the ambiguity in the void definition. Also, \cite{Nadathur2015} argued that the properties of the very largest and deepest voids, i.e. the ones of greatest interest and also the most likely to undergo merging, are very sensitive to the details of the merging criteria chosen. Nevertheless, the environment of voids is a relevant property that influences the gravitational potential ($\Phi$) and therefore the ISW signal due to its time-dependence ($\dot{\Phi}$). Along these lines, \cite{Nadathur2015} found a hint that under-compensated voids with volume-weighted average density $\bar{\delta_{\rm g}} < 0$ might tend to cluster together in space, with
\ba
\bar{\delta_{\rm g}} = \frac{1}{\bar{\rho}} \frac{\sum_{i} \rho_{i} V_{i}}{\sum_{i} V_{i}}-1
\ea
where $\bar{\rho}$ is the mean tracer density ($\Phi=0$ value for the potential) in the galaxy catalogue and the sum runs over all Voronoi cells that make up the void volume (see Figure 1). Further, they found a simple linear relation between $\bar{\delta_{\rm g}}$ and the local density environment,
\ba
\Delta(R) = \frac{3}{R^{3}}\int_{0}^{R}\delta(r)r^{2}{\rm d}r,
\ea
that effectively determines the value of $\Phi$. In this framework, $\Delta(R=3\tilde{R}_{v})<0$ suggests that the isolated void is not compensated by surrounding high density regions. Such voids will correspond to regions of $\Phi>0$, i.e. $\Delta T_\rmn{ISW}<0$ assuming decaying gravitational potentials and standard perturbation theory in $\Lambda$CDM models. With this cut, voids-in-clouds can effectively be removed.

This proxy for the $\Phi$ potential offers a good chance to refine the selection of under-compensated voids, potentially better than void size alone. However, accurate relations are needed for a detailed study of the underlying gravitational potential that might depend on the shape of the density profile and the size of voids, as noted by \cite{Nadathur2015}. Such a relation has actually been found by \cite{NadathurEtal2016} who considered an observable quantity $\lambda$ that is tightly correlated with $\Phi$, namely
\ba
\lambda\equiv\overline\delta_g\left(\frac{R_{\rm eff}}{1\;h^{-1}\rmn{Mpc}}\right)^{1.2}
\ea
where $R_{\rm eff}$ is the effective radius of voids. The majority of voids identified by {\small ZOBOV} correspond to local underdensities within globally overdense regions and thus do not give $\Delta T_\rmn{ISW}<0$. Applying a selection cut $\lambda<0$ selects those globally undercompensated voids which correspond to regions with $\Phi>0$ and thus a negative ISW shift, again assuming decaying gravitational potentials in $\Lambda$CDM perturbation theory.

\subsection{Minimal voids $\sim$ \emph{super}voids}

The analysis of isolated voids offers a great way to probe the ISW effect. However, the combination of their definition and the observed clustering of $\bar{\delta_{\rm g}}<0$ voids suggests that in fact many of them are imbedded in more extended background underdensities, or \emph{super}voids. The largest minimal voids offer an interesting possibility to probe the ISW imprint of these super-structures, i.e. the largest fluctuations in $\Phi$. We also expect that void depth is important, as expressed in Eq.~(3), and centres of isolated voids with the lowest $\lambda$ values are the centres of the largest surrounding \emph{super}voids.

We start with a catalog of $6565$ minimal \texttt{ZOBOV} voids using both CMASS and LOWZ spec-z data from BOSS DR12, spanning $0.15<z<0.7$ in redshift. The $\lambda\sim\Phi$ relation has not been tested on minimal voids, therefore we cannot use it blindly to bin our data. We instead follow the simple and more approximate prescription by \cite{Nadathur2015} and select voids with weighted average densities $\bar{\delta_{\rm g}}<0$ or equivalently $n_{\rm avg}/\bar{n}<1$ (the average tracer number density within voids compared to the mean tracer density). This cut guarantees a void population of $\Phi>0$ with $\Delta T_\rmn{ISW}<0$. We further prune the void catalogue by removing objects possibly affected by failures of \texttt{ZOBOV}'s tessellation method\footnote{\texttt{EdgeFlag}<2 applied to \texttt{ZOBOV}'s corresponding parameter}, leaving $1446$ voids out of $6565$. In Figure 2, we show a summary of selected void parameters for our pruned catalogue including radius, central underdensity, redshift, and angular size. A comparison to Gr08 \emph{super}voids indicates that we are in good position to probe the claims by Gr08 using more large voids to potentially illuminate a true signal, and with a large amount of small voids to see how such signals might disappear.

We thus apply an aggressive cut that removes $\approx78\%$ of the total catalogue. The full BOSS DR12 isolated void catalogue\footnote{The DR12 minimal catalogue we use is non-public but a public catalogue of minimal voids is available for BOSS DR11 at \texttt{http://www.icg.port.ac.uk/stable/nadathur/voids/}}, used as a base data set for the pruned analysis presented in \cite{NadathurCrittenden2016}, contains 10492 voids. For our main conclusions, we will apply additional motivated pruning to the minimal void sample, leaving only the 96 largest minimal voids for the fiducial stacking analysis. For comparison, $1392$ isolated voids are imbedded in these 96 \emph{super}voids while \cite{NadathurCrittenden2016} used $2445$ isolated voids.

\begin{figure*}
\begin{center}
\includegraphics[width=172mm]{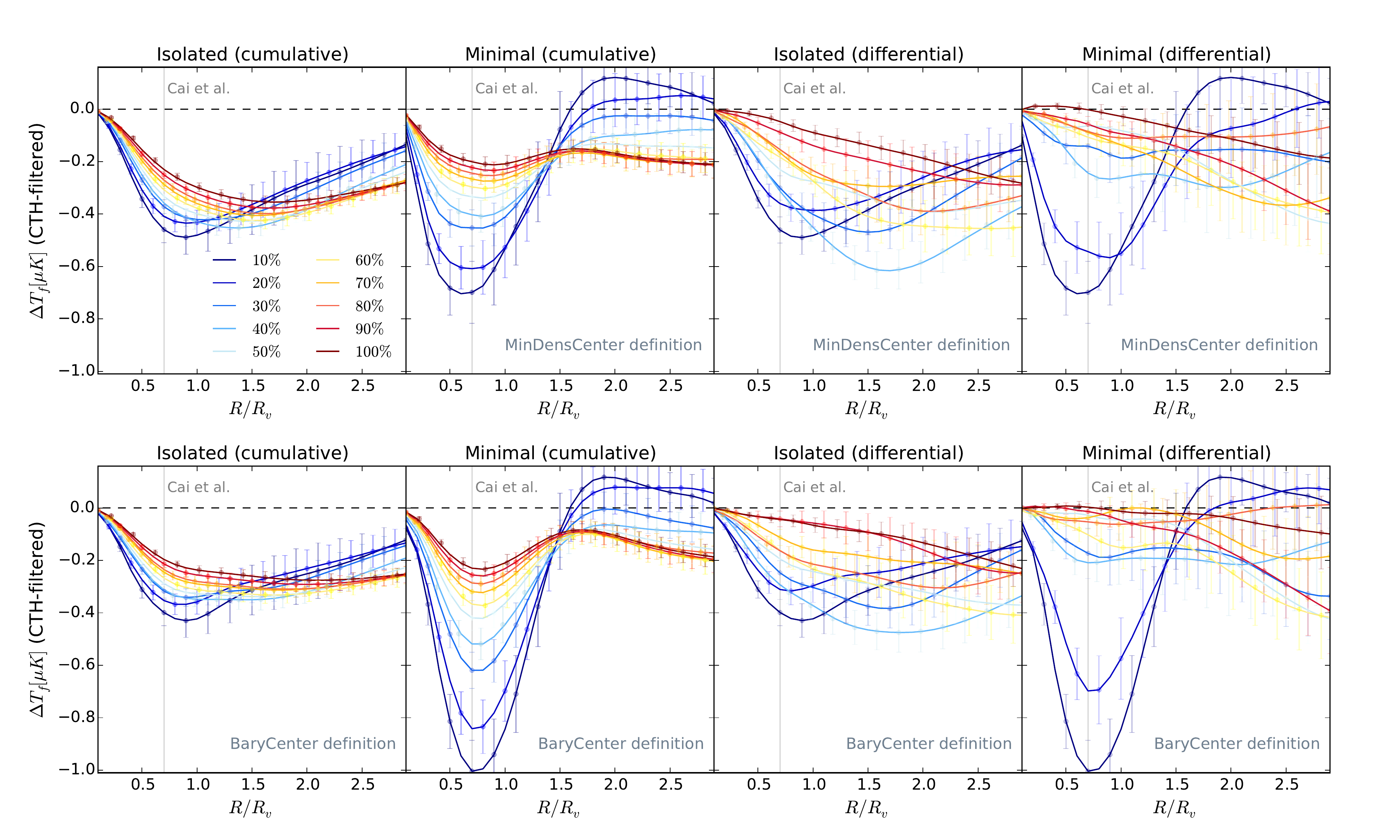}
\caption{A color-coded set of curves shows the cumulatively or differentially binned CTH-filtered ISW profiles. We compare isolated and minimal voids with both stacking protocols. Subfigures in the top row show our results using MinDensCenter definition while the bottom panel corresponds to BaryCenter void centering. The vertical line indicates the best re-scaling value obtained by Cai et al. (2016) for their void sample. For this test, we removed the $2\leq \ell \leq10$ modes from the ISW map.}
\end{center}
\end{figure*}

We note that not all the minimal voids in the catalogue have actually been merged, as demonstrated in Figure 2. A parameter $N_{\rm merged}$ indicates the number of isolated sub-voids, reaching $N_{\rm merged}\approx40$ values for the largest \emph{super}voids. Without an actual cut applied to the data, we mark objects with $N_{\rm merged}\geq5$ values and see a good correlation with a subset defined by the often considered $p_{\rm void}>3\sigma$ probability cut (see e.g. \cite{ZOBOV} for details) that selects the largest and deepest objects in the sample. 

Nevertheless, we are guided to use simulations to {\it a priori} decide exactly which minimal voids to stack for detecting a specific ISW signal of \emph{super}voids.

\subsection{The Jubilee simulation}

We first analyzed simulated data from the Jubilee ISW project \citep{Watson2014} to estimate the $\Lambda$CDM expectation for the stacked ISW imprint of voids, following \cite{Hotchkiss2015}. The Jubilee ISW project is built upon the Jubilee simulation, a $\Lambda$CDM N-body simulation with $6000^3$ particles in a volume of ($6h^{-1}$ Gpc)$^{3}$, assuming WMAP-5 cosmology. A corresponding mock LRG catalogue was initially designed to model the properties of SDSS LRGs studied in \cite{Eisenstein2005}. This mock provides a sample with $\bar{n}\approx8\times10^{-5}h^{3}$ Mpc$^{-3}$ that is slightly lower than the corresponding BOSS values. For CMASS the co-moving number density peaks at $z\approx0.5$ with $\bar{n}\approx4\times10^{-4}h^{3}$ Mpc$^{-3}$, while it is $\bar{n}\approx3\times10^{-4}h^{3}$ Mpc$^{-3}$ for LOWZ that slightly depends on redshift. 

This difference could affect our conclusions about the optimal stacking strategy. In sparser galaxy tracers, the number of voids identified decreases, but the average void size is larger \citep{Sutter2014_DM,Nadathur2015}. More importantly, voids resolved by sparse galaxy samples also on average trace shallower but more extended dark matter underdensities \citep{Nadathur2015}, which should have a longer photon travel time and therefore correspond to larger ISW temperature shifts. This conclusion is validated by the results of \cite{Hotchkiss2015}. They estimated the ISW imprint for voids (and also superclusters) in mock LRG catalogues with differing brightness and sparsity in the Jubilee simulation. They found that the sparser sample gave consistently larger ($\approx30\%$ at the peak) $|\Delta T|$ with similar shapes for the CTH-filtered signals in most of the profile. This suggests that the difference is certainly below the level of CMB noise in the measurement for the galaxy number densities given in our analysis. We therefore conclude that the expected stacked ISW signal we determine from Jubilee will be an \emph{overestimate} of that observable from superstructures in the BOSS data.

We make use of two catalogues of voids in the Jubilee LRG mock using the \texttt{ZOBOV} algorithm. We considered the full-sky LRG mock data set with the LOWZ and CMASS redshift windows. The mock catalogue of isolated Jubilee voids consists of $19528$ voids. As expected, the minimal version is less numerous with $11043$ objects after sub-void merging. The additional pruning cuts we described in Section 3.2 result in $2617$ minimal voids and $7446$ isolated voids with properties $\bar{\delta_{\rm g}}<0$ and \texttt{EdgeFlag}<2.

\section{Jubilee ISW analysis}

Following \cite{Hotchkiss2015} and \cite{Kovacs2016}, we stack the ISW-{\it only} Jubilee temperature map on void locations. We re-scale the images knowing the angular size of voids. On the stacked images, we then measure azimuthally averaged radial ISW profiles in $R/R_{v}$ fractional void radius units. While environment, density profiles, redshifts, and exact shapes can be important for the accurate estimates, the ISW signal is expected to correlate with void size. In our fiducial sample, therefore, we order the voids by their $\tilde{R}_{v}$ radius. We also split and explore our data in the following ways:
\begin{itemize}
\item most importantly, we compare the imprints of isolated and minimal voids in all possible aspects.
\item secondly, we measure non-filtered and CTH-filtered profiles for the objects and make a comparison.
\item based on the $\tilde{R}_{v}$ effective radius parameter of voids, we create 10 bins for $10\%$ percentiles.
\item using the binned data, we stack the images both cumulatively and differentially.
\item we test the importance of the void center definition; barycenter vs. minimum density center.
\item we probe the effects of removing the $2\leq \ell \leq10$ large-scale modes from the ISW map.
\item beyond the fiducial ascending ordering based on $\tilde{R}_{v}$, we try alternative orderings based on $\lambda$, $N_{\rm merged}$, and $p_{\rm void}$.
\item we show how combined modifications affect the results.
\end{itemize}

For better insights into $S/N$ properties of these very different measurement configurations, we estimated statistical uncertainties by stacking the Jubilee ISW temperature map on random positions (see below figures 3, 4, and 5).

\begin{figure*}
\begin{center}
\includegraphics[width=178mm]{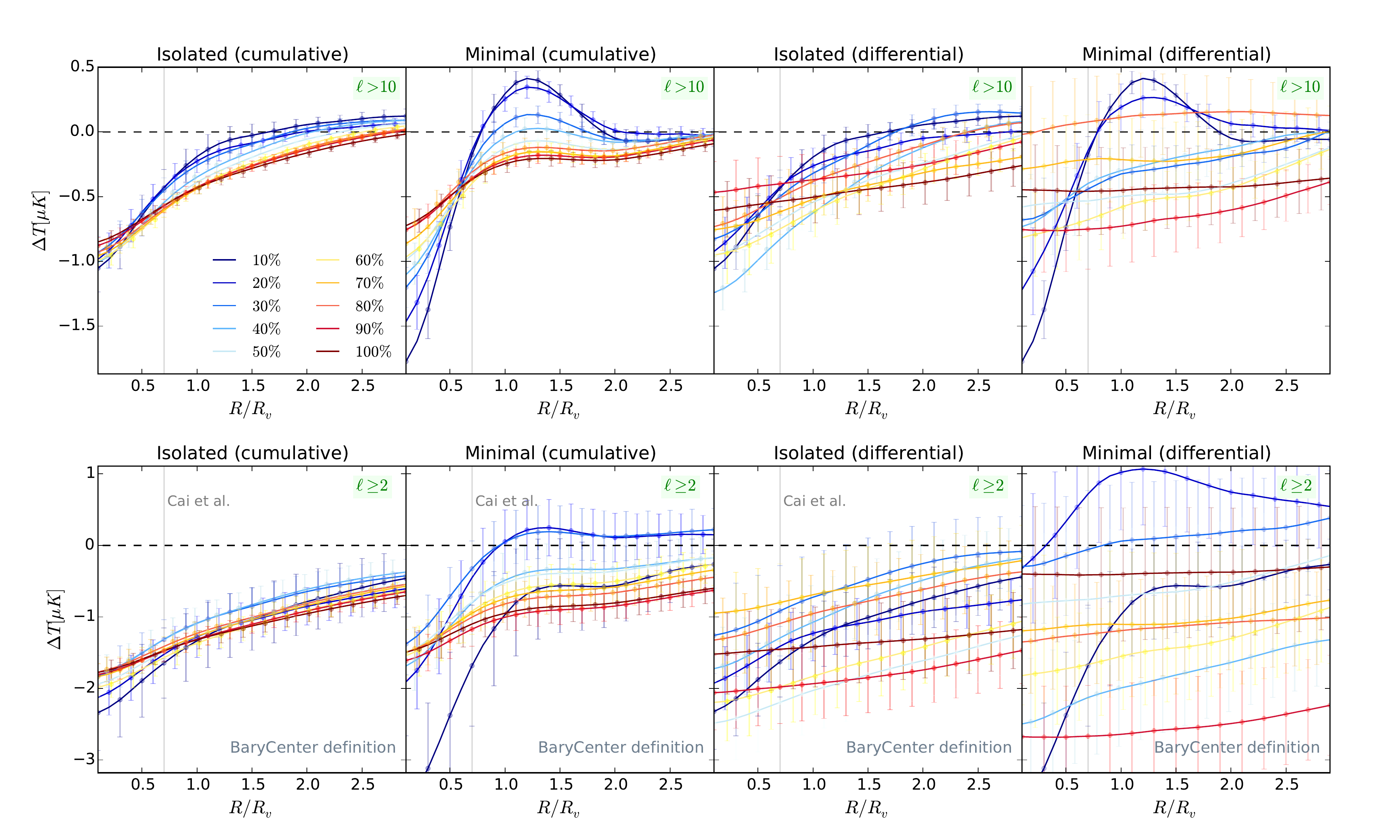}
\includegraphics[width=178mm]{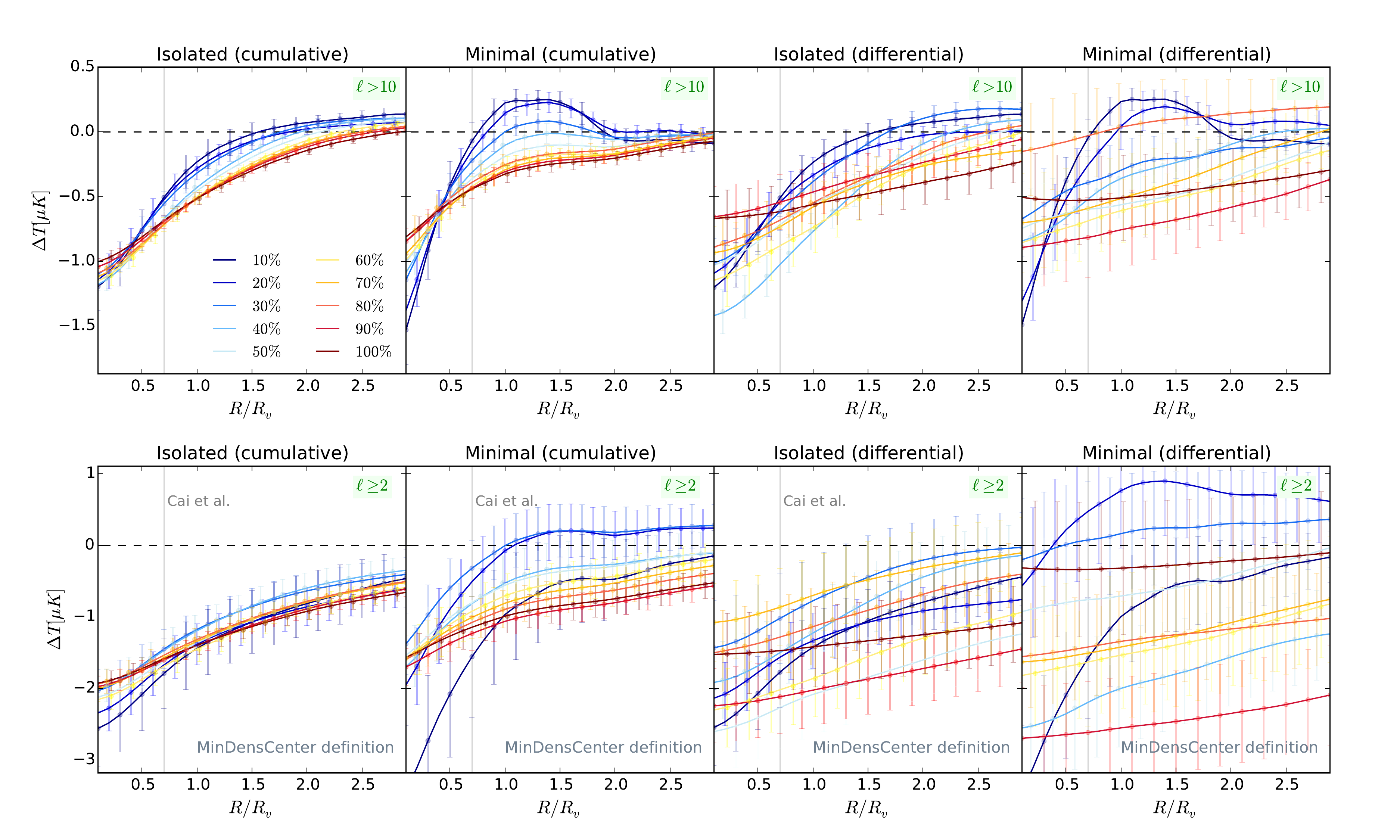}
\caption{Azimuthally averaged ISW imprint profiles of different voids are compared as a function of $R/R_{v}$. Subfigure labels indicate the stacking protocol, void type, and void center definition. The role of the large-scale $2\leq \ell \leq10$ modes in reducing the fluctuations and biases may be investigated by comparing the first and second rows of subfigures for BaryCenter definition, and by comparing the third and fourth rows of subfigures for MinDensCenter definition.}
\end{center}
\end{figure*}

\subsection{CTH filters}

In Figure 3, we start our analysis by showing how isolated and minimal voids compare in the most traditional CTH-filtering scheme with filtered CMB temperatures 
\ba
\Delta T_{f} = \frac{\int_{0}^{R}\Delta T(r)dr}{\int_{0}^{R}dr} - \frac{\int_{R}^{\sqrt{2}R}\Delta T(r)dr}{\int_{R}^{\sqrt{2}R}dr}
\ea
where $R$ is the filter radius. Centered on the voids, CMB temperatures are averaged within a circular aperture $r < R$, and then the background temperature averaged in an equal-area concentric annulus with $R < r < \sqrt{2} R$ is subtracted in order to measure the possible ISW imprints. For minimal voids, we can immediately validate the approximate findings by \cite{Cai2016} who reported in their simulations that a re-scaling of CTH filters with a $R/R_{v}\approx0.7$ factor guarantees the highest filtered signal (second panels from the left). On the other hand, figure 3 is also helpful to validate the findings by \cite{Hotchkiss2015} who found that this property does depend on the void definition; the cumulative stacking signal of isolated voids peaks closer to $R/R_{v}\approx1$. 

Apart from the different peak location for the CTH-filtered signals, the shape of the signal appears to change much less for isolated voids as, cumulatively, smaller voids are added to the stacking. The differential binning of voids is helpful to understand this property even if their corresponding error bars are naturally bigger compared to the cumulative case (see the right panels of figure 3). For isolated voids, the filtered signals tend to show a peak at $R/R_{v}\approx1$ as in the cumulative case but then the location of this peak shifts to larger $R/R_{v}$ values with smaller and smaller voids in the stack. The smaller half of the isolated sample ($\geq50\%$) appears to show a peak at $R/R_{v}\geq2$ indicating that these voids are indeed typically imbedded in larger cold ISW spots that are not unique to them. These findings hold for both BaryCenter and MinDensCenter (the center of the largest empty sphere that can be inscribed within the void) void center definitions. We conclude here that although the precision improves greatly as error bars shrink when using $\sim100\%$ of the data, differential stacking results show very different profiles for different void size bins. Therefore, a full combination of all voids in one stack appears to be suboptimal.

On the other hand, minimal voids show different ISW imprints in Jubilee. The rightmost panel in figure 3 shows that the largest $\approx20\%$ of the sample produces a peak in the CTH-filtered signal at $R/R_{v}\approx0.7$. In fact this largest $\approx20\%$ part of the minimal data appears to dominate the cumulative stacking results while the rest of the sample behaves similarly to the isolated case. This feature reflects the importance of void merging because these are typically non-merged voids in the minimal catalogue.

An interesting feature in the data is the stronger imprint for this top $\approx20\%$ population when considering BaryCenter definition, because a void's deepest region (MinDensCenter) is expected to correspond best to the peak ISW signal \citep[see discussions by][]{NadathurEtal2016}. We note that for isolated voids this expectation appears to be true. Among other properties, we investigate this difference in greater details in our additional tests below.

\subsection{Non-filtered profiles and large-scale modes}

We then consider azimuthally averaged radial $\Delta T (R/R_{v})$ profiles without CTH filtering. We later use these profiles as {\it templates} for the expected ISW profile of BOSS voids, effectively defining a hybrid approach; in between the CTH measurement by \cite{Cai2016} and the optimal matched filtering technique by \cite{NadathurCrittenden2016}. 

In Figure 4, we show how these profiles compare for size-ranked and binned isolated and minimal voids. We observe that the magnitude of the central ISW signal is comparable but the shape of the imprint in $R/R_{v}$ units is different. When combining all data in the stacking, the more numerous isolated voids are expected to have higher $S/N$, with presumably higher covariance, even if the minimal voids have larger typical angular size that means smaller CMB fluctuations. 

We also compare the signals with and without the $2\leq \ell \leq10$ modes in the ISW-only map. Without these modes, we observe a reduced noise and no significant bias in the measured profiles, as expected, at the expense of removing some of the ISW signal. Relevantly, the use of these large-scale fluctuations is yet another difference between the two recent BOSS DR12 analyses; \cite{Cai2016} removed the $2\leq \ell \leq10$ modes while \cite{NadathurCrittenden2016} used all available modes. With our simulation tests, we can validate both choices to some extent. Firstly, the cumulative stacking of isolated voids with all $\ell \geq 2$ modes included shows a factor of $\approx2$ stronger imprints in the central region compared to the $2\leq \ell \leq10$ case (leftmost panels in Figure 4). In both cases, the imprints are significantly more extended than the void radius, reaching $R/R_{v}\gsim3$. However, the differential binning scheme shows large fluctuations compared to the errors and no clear trend in the stacked ISW profiles. These features are introduced by the large-scale modes in the Jubilee ISW map, highlighting the importance of cosmic variance. See second sub-plots from the right in the top rows of Figure 4.

The behavior of the largest $\approx20\%$ of minimal voids is different, as shown in the rightmost sub-figures in Figure 4. The differential binning technique proves that the shape of their imprint profile in $R/R_{v}$ units is more compact than that of the rest of the sample. With $\ell \geq10$ modes, this qualitative difference becomes clear with cold regions close to the re-scaled void centers and hot imprints in the surroundings. However, zero crossings are seen at slightly different locations for BaryCenter and MinDensCenter definitions and also the shape of the inner profiles seems to be different. With $\ell \geq2$ modes, however, the ISW imprints significantly fluctuate and show biases in the radial temperature profile even for the cumulatively stacked sample but importantly for the differential binning scheme. 

Given the measurement errors, we thus conclude that the removal of the $2\leq \ell \leq10$ modes helps to remove potential biases from the measured profiles, resulting in a convergence to zero signal at $R/R_{v}\approx2$ for these largest voids (see the first two rows of Figure 4). The rather special ``cold spot plus hot ring" ISW imprint of the largest $\approx20\%$ of minimal voids becomes more easily testable. 

\begin{figure*}
\begin{center}
\includegraphics[width=178mm]{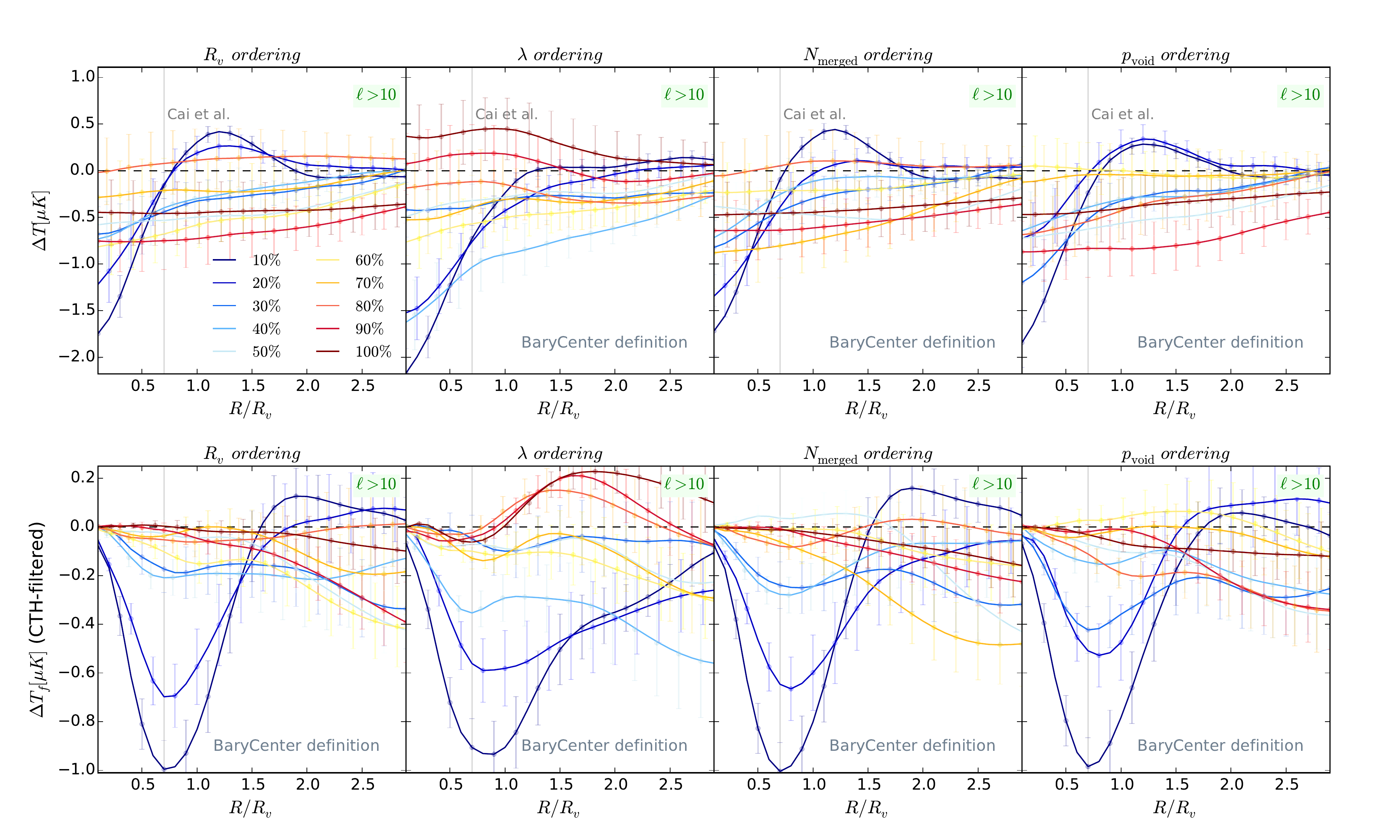}
\includegraphics[width=178mm]{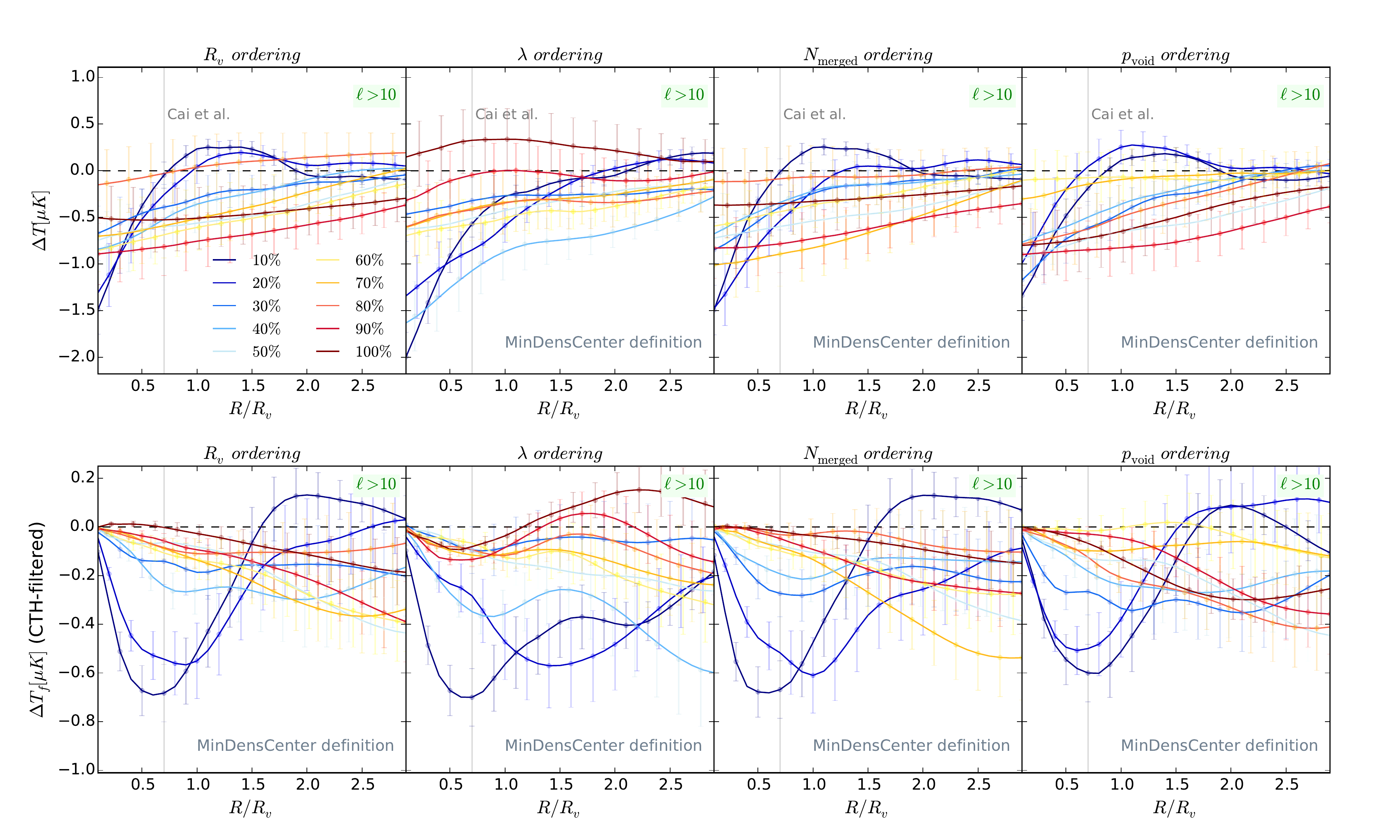}
\caption{Ordering strategies are compared for BaryCenter (top two rows) and MinDensCenter definitions (lower two rows). Beyond the fiducial $\tilde{R}_{v}$ ordering, we considered other schemes based on parameters $\lambda$, $N_{\rm merged}$, and $p_{\rm void}$. Color-coded curves show differentially stacked results in the usual $10\%$ bins. First and third rows show the results without filtering as a function of $R/R_{v}$, while sub-figures in the second and fourth rows correspond to CTH-filtered profiles. The top $\approx20\%$ of the data is qualitatively different in each case and a good overall consistency is seen in this test. }
\end{center}
\end{figure*}

\subsection{The role of void ordering}

As a final part in the Jubilee stacking analysis, we aim to test the importance of the ordering scheme applied to the data. We considered the $\ell \geq 10$ ISW map for these tests. In Figure 5, we demonstrate, for both CTH-filtered and non-filtered profiles, that the special nature of the largest $\approx20\%$ of minimal voids is observable not just when the fiducial $\tilde{R}_{v}$ based ordering is applied, but also for $\lambda$, $N_{\rm merged}$, and $p_{\rm void}$ orderings. This is not surprising because these void parameters are expected to correlate, as seen in Figure 2 for the BOSS DR12 data set. As a consequence, the top $\approx20\%$ cut applying ordering based on the number of merged sub-voids corresponds to $N_{\rm merged}\gsim5$ (no actual cut was applied).

We also note that the $\lambda$ relation has not been calibrated for minimal voids but the $\lambda$ ordering appears to perform well; the most ISW-sensitive voids produce the coldest imprints in the void centre. However, voids with $\lambda$ values approaching zero appear to leave a $\Delta T_\rmn{ISW}>0$ imprint that points to the need of proper calibration of this technique.

Finally, we again observe differences between stackings using center definitions. The signal is typically suppressed for BaryCenter definition, as expected, but for the top $\approx20\%$ the signal is slightly increased. This feature in the data appears to be less outstanding in simple radial profiles than with CTH-filtering. We note that since these void center definitions are not expected to differ significantly, the importance and origin of this $\sim1-2\sigma$ feature is unclear at the moment, but a possible cause is simply cosmic variance.

\subsection{ISW template for \emph{super}voids}

The hot ring feature around the centers of the largest $\approx20\%$ of minimal voids suggests that these objects are good candidates to be called \emph{super}voids because this additional ISW feature is expected to be caused by the neighboring superclusters, i.e. neighbors in the supercluster-\emph{super}void network in the cosmic web \citep[see e.g.][]{Einasto1997}. Most importantly, this property makes these BOSS \emph{super}voids more similar to the super-structures seen in SDSS and DES data that have shown anomalously high ISW-like signals.

A natural argument against this conclusion is that the stacked signal of isolated voids with $\ell \geq 2$ modes is higher in magnitude. The estimated $S/N$ is also higher because there are more isolated voids for a given physical volume and tracer catalogue. However, the $R/R_{v}\gsim3$ extent of the imprints of isolated voids suggests that these often small voids are imbedded in more extended background underdensities or \emph{super}voids. The Jubilee simulation showed that cold spots associated with deep isolated voids are in fact not unique to them but produced due to the evolution of larger fluctuations in the gravitational potential. This might lead to a significant covariance between imprints of isolated voids; a feature that has in fact been reported by \cite{NadathurCrittenden2016}. Naturally, we also expect a non-negligible covariance for the \emph{super}void sample due to overlap effects that we will take into account in the analysis.

\section{BOSS ISW analysis}

We have characterized the resulting shape and amplitude of the ISW imprints for different void definitions in the Jubilee simulation. We now perform measurements with BOSS minimal voids using the {\it a priori} selected measurement parameters. Note that we have advanced the CTH methodology by producing template profiles based on the Jubilee stacking of different void types. However, we also show results using the traditional CTH filters for completeness. We define our data set as follows.

We again note that galaxy sampling rates and thus void size distributions are different in Jubilee and BOSS. Galaxy voids might be imbedded in even slightly larger but shallower dark matter underdensities, considering the discussion in Section 3.3. Therefore, fractional void size bins do not exactly contain voids of the same physical size for the data and for the simulation. In our Jubilee mock minimal void catalogue, the largest $\approx20\%$ with the fiducial $\tilde{R}_{v}$ ordering defines a set of voids with $\tilde{R}_{v}\gsim110~Mpc/h$. Therefore, we are led to conservatively select the minimal voids of the same physical size in our BOSS analysis with $\tilde{R}_{v}\gsim110~Mpc/h$ for a safe analysis of truly the largest objects, i.e. the ones which showed cold spot plus hot ring imprints in Jubilee.

This final data set contains $96$ \emph{super}voids that are expected to represent the largest half of the $p_{\rm void}>3\sigma$ subset shown in Figure 2. In total, these \emph{super}voids contain $1392$ isolated sub-voids with $5\leq N_{\rm merged}\leq47$.

\subsection{CMB temperature data}
For our cross-correlations, we use the {\it Planck} Spectral Matching Independent Component Analysis (SMICA) CMB temperature map \citep{Planck_15} downgraded to $N_{side}=512$ resolution with \texttt{HEALPix} pixelization \citep{healpix}. We mask contaminated pixels with the $N_{side}=512$ WMAP 9-year extended temperature analysis mask \citep{WMAP9} to avoid re-pixelization effects of the $N_{side}=2048$ CMB masks provided by {\it Planck}. Several studies confirmed \citep[see e.g.][]{Planck19} that the cross-correlation signal detected at void locations is independent of the CMB data set when looking at WMAP Q, V, W, or {\it Planck} temperature maps. We, however, again checked for possible color dependence in the analysis.

\begin{figure}
\begin{center}
\includegraphics[width=83mm]{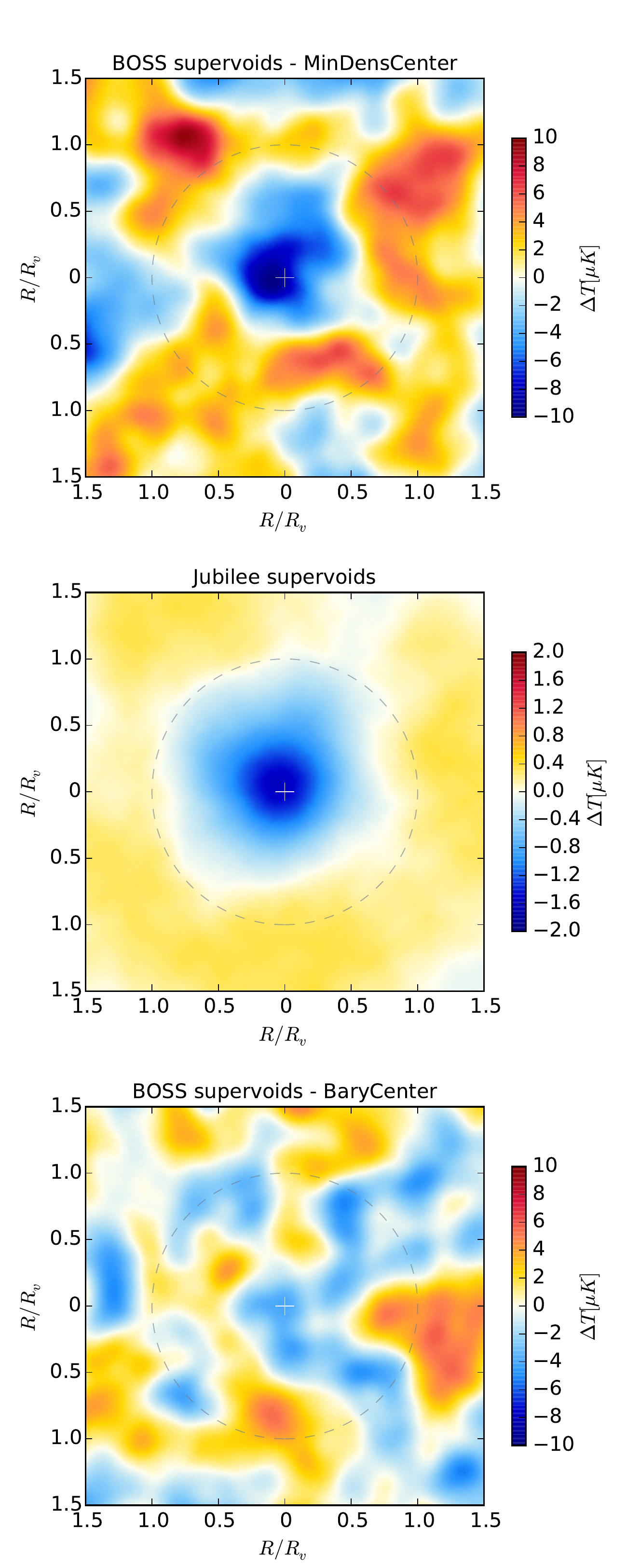}
\caption{Imprint of \emph{super}voids in BOSS data and in the Jubilee simulation. For BOSS data, we applied a smoothing to the individual raw CMB images only for this illustration using $\sigma=3^{\circ}$ symmetrical Gaussian beam in \texttt{HEALPix}. The data shows higher-than-expected imprints for MinDensCenter definition but appears to be rather normal for BaryCenter defined voids. Mind the $5\times$ smaller temperature for the simulated imprint. The dashed circle marks the void radius at $R/R_{v}=1$.}
\end{center}
\end{figure}

\begin{figure*}
\begin{center}
\includegraphics[width=83mm]{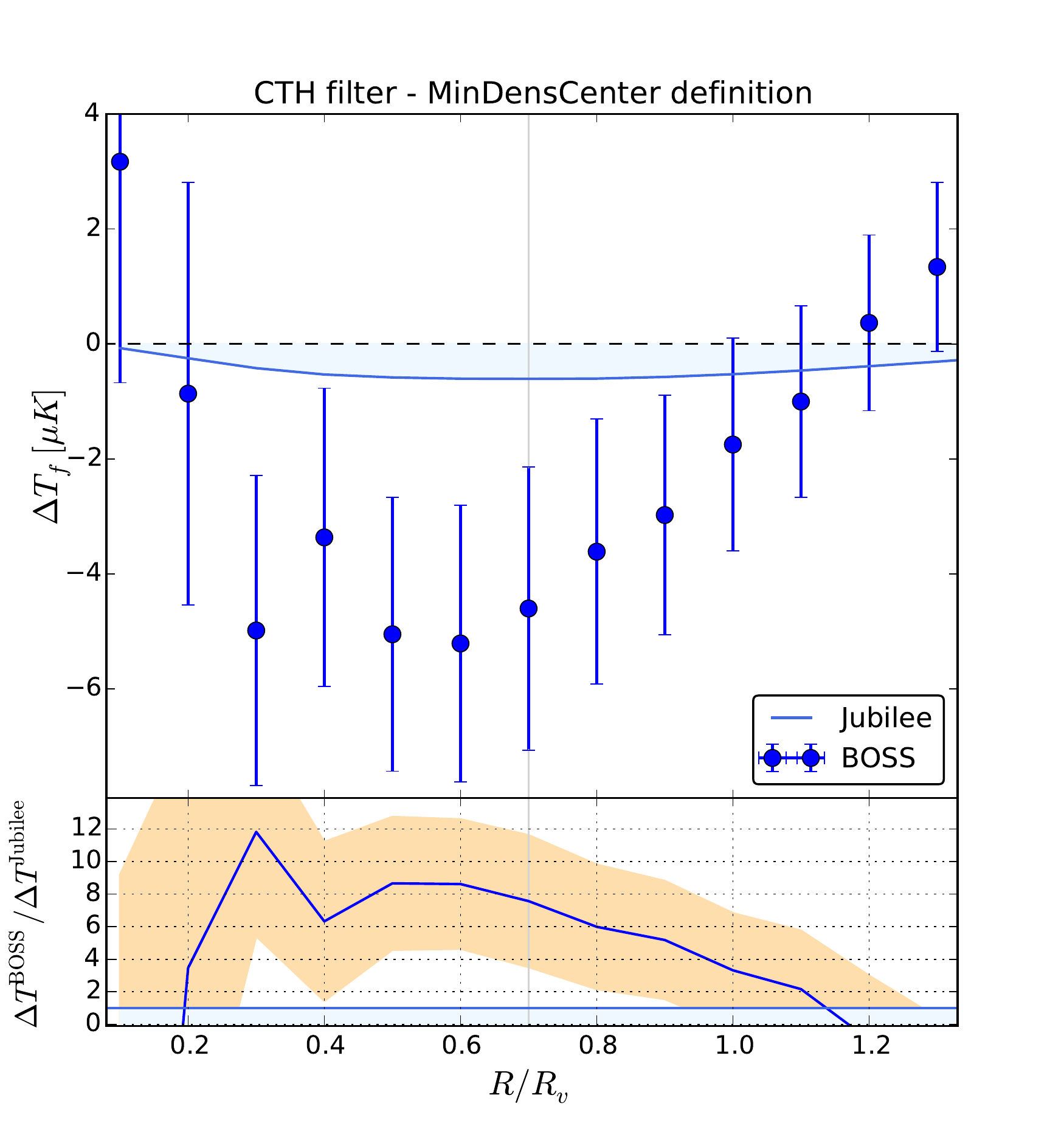}
\includegraphics[width=83mm]{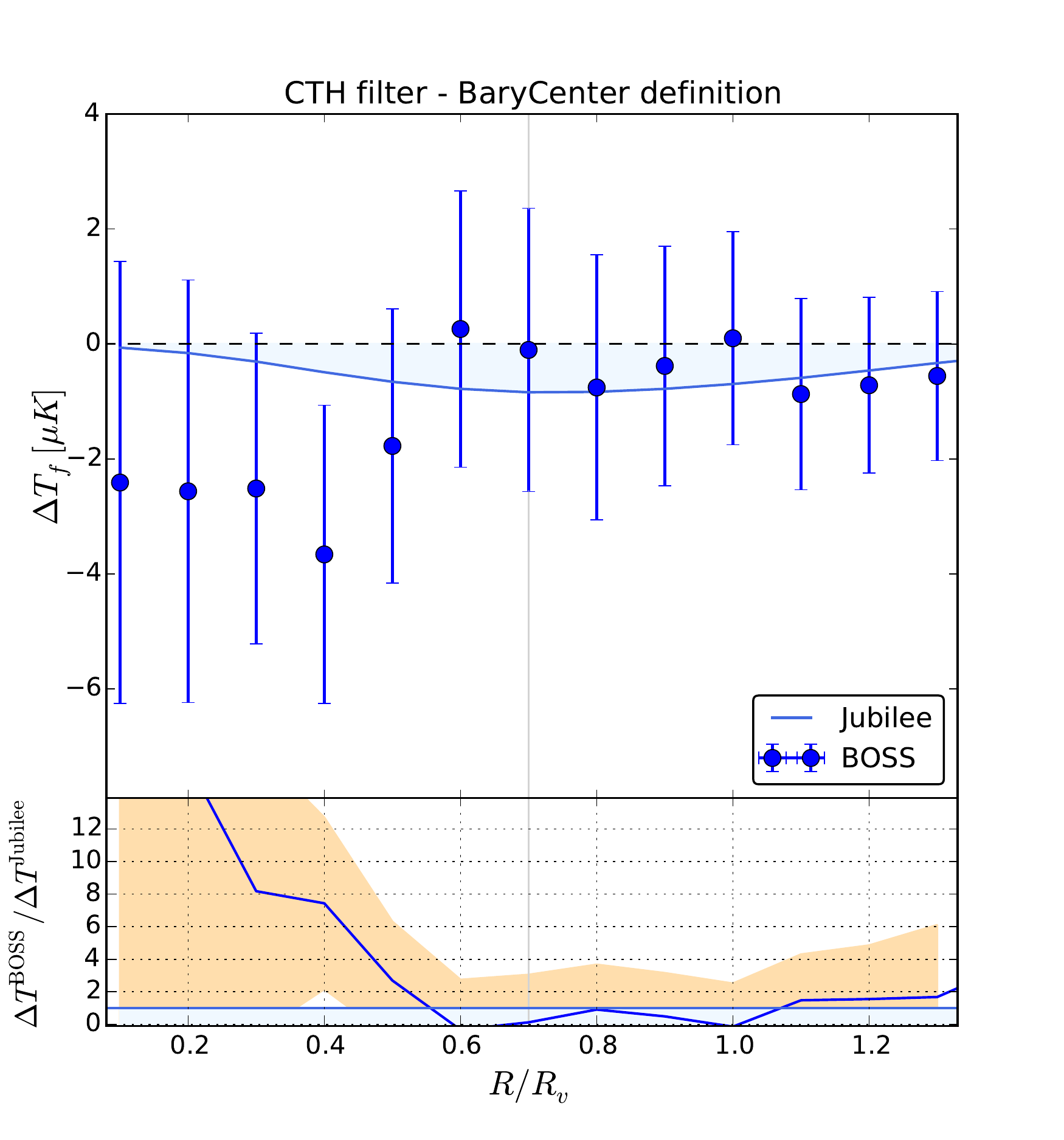}
\caption{CTH-filtered temperatures profiles of BOSS \emph{super}voids are compared to those of Jubilee \emph{super}voids. Results using BaryCenter and MinDensCenter definitions are compared, including point-by-point $\Delta T^\rmn{BOSS}/\Delta T^\rmn{Jubilee}$ values shown in the bottom of the sub-figures. The vertical line marks the $R/R_{v}\approx0.7$ CTH re-scaling parameter where the filtered signal peaks in the simulation analysis. The absolute error bars for $\Delta T^\rmn{BOSS}$ in top panels and the errors relative to $\Delta T^\rmn{Jubilee}$ in the bottom panels (marked by the yellow shaded area) are based on the $1000$ random stacking measurements using Gaussian CMB simulations that we describe in Section 5.3.}
\end{center}
\end{figure*}

\subsection{Stacked images of BOSS \emph{super}voids}

We then create a stacked image of the 96 BOSS \emph{super}voids using the {\it Planck} SMICA CMB temperature map. As in the simulation analyses, we remove $\ell<10$ modes and re-scale each image around the void center. The stacked signal in $R/R_{v}$ units is shown in Figure 6. We perform the stacking using both BaryCenter and MinDensCenter definitions and compare the results to the corresponding Jubilee image. 

For the MinDensCenter case, the BOSS data shows a visually compelling $\Delta T \approx -10 ~\mu K$ cold imprint in the central region of the image at $R/R_{v}\lsim0.7$ and a $\Delta T > 0$ area in the surroundings. In its nature, this imprint appears to be very similar to the Jubilee result, with the BOSS imprint being more compact and having higher amplitude. 

The amplitude of the imprint appears to be more modest for the BaryCenter definition but the shape is again similar. This reduced imprint is not unexpected since the deepest regions of voids are expected to correspond to the coldest ISW imprints, even if in our simulation we found a different trend for some voids, possibly due to cosmic variance.

\begin{figure*}
\begin{center}
\includegraphics[width=83mm]{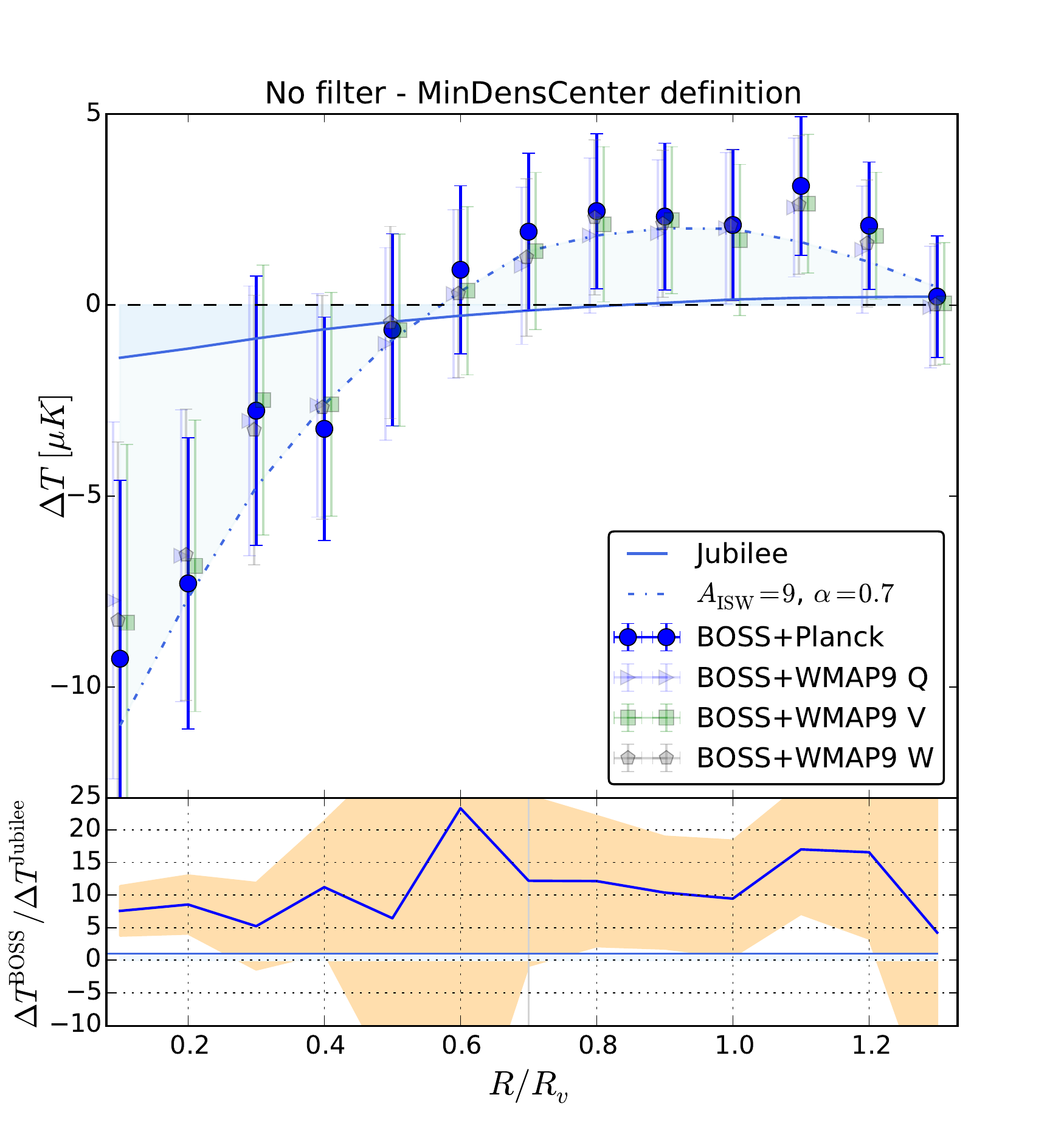}
\includegraphics[width=83mm]{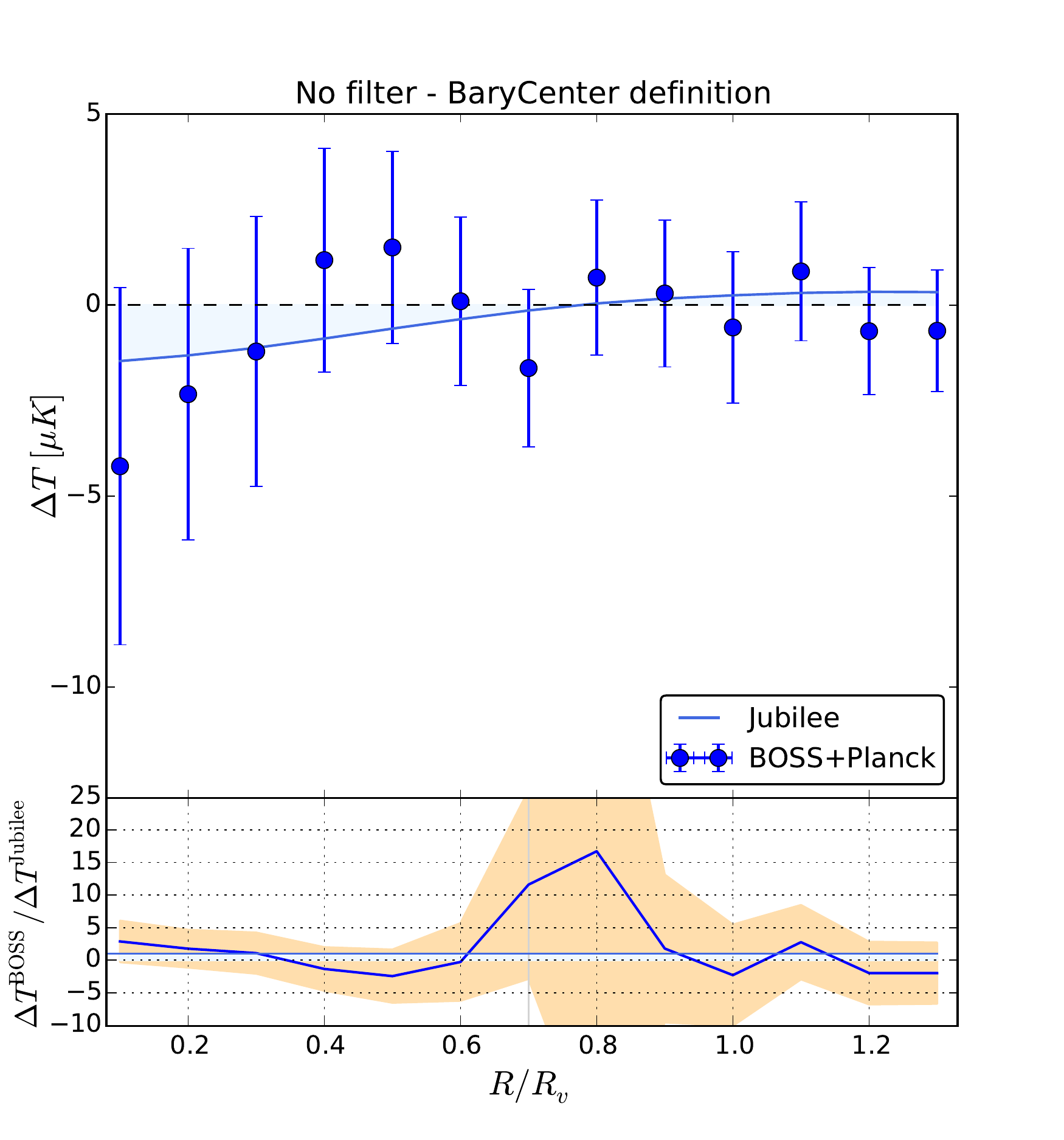}
\caption{Template fitting results of BOSS \emph{super}voids are compared to those of Jubilee \emph{super}voids. Results using BaryCenter and MinDensCenter definitions are compared, including point-by-point $\Delta T^\rmn{BOSS}/\Delta T^\rmn{Jubilee}$ values shown in the bottom of the sub-figures. We find evidences for a rather high $A_{\rm ISW}\approx9$ amplitude plus a need for an additional $\alpha \approx0.7$ radial re-scaling parameter to match the observation with the template profile. The absolute error bars for $\Delta T^\rmn{BOSS}$ in top panels and the errors relative to $\Delta T^\rmn{Jubilee}$ in the bottom panels (marked by the yellow shaded area) are based on the $1000$ random stacking measurements using Gaussian CMB simulations that we describe in Section 5.3.}
\end{center}
\end{figure*}

\subsection{CTH filter analysis}

We then measure the traditional CTH-filtered CMB temperatures as a function of $R/R_{v}$ filter re-scaling to quantify the results. We estimated statistical errors by performing $1000$ random stacking measurements using Gaussian CMB simulations. The randoms have been generated with the \texttt{HEALPix} \citep{healpix} \texttt{synfast} routine using the {\it Planck} 2015 data release best fit CMB power spectrum \citep{Planck_15}. Gaussian CMB simulations without instrumental noise suffice because the CMB error is dominated by cosmic variance on the scales we consider \citep[see][]{Hotchkiss2015}. We decided to keep the void positions fixed and vary the CMB realization, because in this case overlap-effects for voids are accounted for more efficiently.

The results, shown in Figure 7, reflect the visual impression. While the signal is mostly consistent with zero for BaryCenter definition, we find evidence for a $\approx2\sigma$ temperature depression for the MinDensCenter case. We measure $A_{\rm ISW}\approx9$  using the presumably optimal $R/R_{v}\approx0.7$ CTH re-scaling parameter, but the filtered signal in fact peaks at smaller radii ($R/R_{v}\approx0.6$).

The signal exceeds the Jubilee expectation and it is comparable to the imprints found by \cite{Kovacs2016} who analyzed \emph{super}voids in the DES footprint. The origin of this difference in the imprints due to different void centre definitions, if understood, might be a key feature to trace the unexpected signals, thus we perform more tests below.

\subsection{ISW template fit analysis}

We fit an $A_{\rm ISW}$ amplitude to the observable imprints in the BOSS data using the ISW template profile we constructed with Jubilee. We evaluate a statistic ${\chi}^2 = \sum_{ij} (\Delta T_i^\rmn{BOSS}-\Delta T_i^\rmn{Jubilee} )C_{ij}^{-1} (\Delta T_j^\rmn{BOSS}-\Delta T_j^\rmn{Jubilee})$ where $C$ is the covariance matrix obtained using the 1000 random stacking measurements. Overall, the BOSS data favors an enhanced amplitude but another unexpected feature is seen in the real-world data; additional re-scaling of radii is needed for a good fit, as shown in Figure 8. As an additional test, we do not find evidence for frequency dependence when using WMAP9 Q, V, and W temperature maps (see left panel of figure 8).

We note that in the case of CTH-filtering the $R/R_{v}\approx0.7$ re-scaling maximizes the filtered signal for the Jubilee \emph{super}voids because of the particular shape of the ISW imprint profiles. In spite of that, the reason for the extra $\alpha$ re-scaling is the more compact ISW-\emph{like} imprint of BOSS \emph{super}voids compared to simulated imprints (see also Figure 6). The origin of this feature remains unclear since it is not motivated by our simulation analysis. 

Possibly, imperfections in void merging algorithms are at fault and the analysis can be improved. We note, however, that DES \emph{super}voids show a similarly dislocated imprint using a different void finder technique \citep{Kovacs2016}, suggesting a real feature in the data.

We characterized this {\it a posteriori} refinement with a $\hat{R}_{v}=\alpha \tilde{R}_{v}$ formula where $\hat{R}_{v}$ is the modified angular extent of the ISW template profile. The best fit value that we find is $\alpha \approx 0.7$. Correlations between radial bins were accounted for using the full covariance that was estimated using the set of CMB simulations described in Section 5.3. The significance of the enhanced $A_{\rm ISW}\approx9$ amplitude is $\approx2.5\sigma$ compared to zero ISW signal, i.e. a $\gsim2\sigma$ discrepancy compared to the Jubilee-based $\Lambda$CDM predictions.

In fact, this pattern offers a qualitative explanation why the ISW-\emph{like} excess signal is greatly reduced considering isolated voids or by introducing slight mis-centering with the BaryCenter definition. In \emph{super}voids, central sub-voids appear to be aligned with extremely cold temperature imprints while outer sub-voids are aligned with $\Delta T>0$ areas, resulting in a peculiar cancellation effect in their jointly stacked imprints; a property that only shows up in real-world data.

This moderately significant observation of excess ISW-like signals in \emph{super}voids is consistent with several previous estimates based, at least in part, on merged voids \citep{GranettEtAl2008,CaiEtAl2014,Cai2016,Kovacs2016}. These observations raise the possibility of a physical connection between the Eridanus \emph{super}void and the CMB Cold Spot but in that case the actual ISW prediction and void parameters are uncertain at the moment.

\subsection{On the Cold Spot - \emph{super}void connection}

We attempt to interpret the case of the Cold Spot as an ISW-\emph{like} effect by using the enhanced $A_{\rm ISW}\approx9$ amplitude and $\alpha\approx0.7$ scaling that we determined empirically in BOSS \emph{super}void data. 

For this exploratory test, we follow \cite{Naidoo2017} and consider the coldest spot in the Jubilee ISW map defined using a Spherical Mexican Hat Wavelet (SMHW) filter of angular size $R_{\rm SMHW}=5^{\circ}$ that was originally used to identify the Cold Spot in the CMB \citep{CruzEtal2004}. The radial ISW-only temperature profile is then measured around the center of the coldest spot, considering the original $\ell\geq2$ Jubilee map and also the version without $\ell<10$ modes used in the main BOSS analysis. 

\begin{figure*}
\begin{center}
\includegraphics[width=179mm]{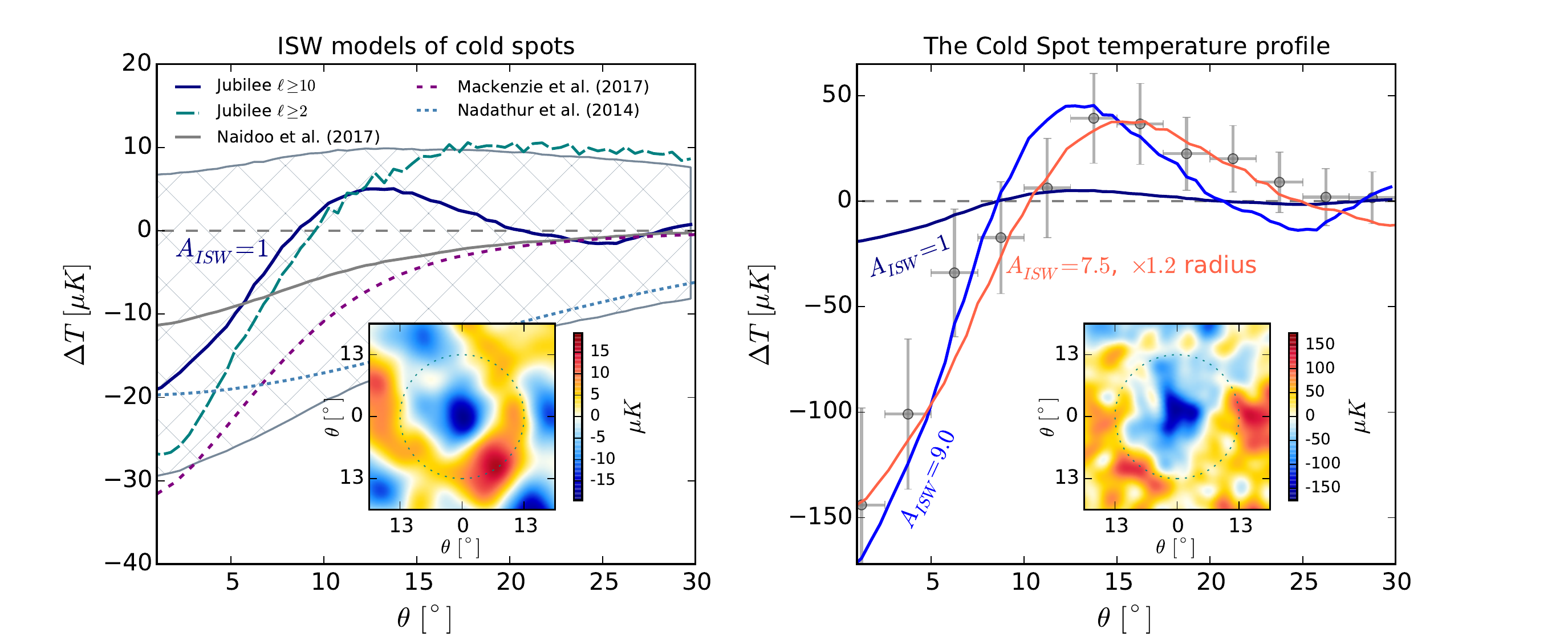}
\caption{Left: ISW profiles of the coldest spot in Jubilee ($R_{\rm SMHW}=5^{\circ}$ filtering) with and without large-scale modes. We compare these profiles with various model estimates for the ISW contribution to the Cold Spot and report good consistency. The inset shows the $\ell\geq10$ Jubilee temperature data in the location of the coldest spot. Right: the $\ell\geq10$ $A_{\rm ISW}\approx1$ Jubilee coldest spot template (dark blue) and an enhanced $A_{\rm ISW}\approx9$ Jubilee coldest spot template (blue) are compared to the Cold Spot data (gray points and errors). Improved agreement is seen if the angular size of the imprint is increased {\it a posteriori} by $20\%$ and if the amplitude is modified to $A_{\rm ISW}\approx7.5$, still in line with BOSS findings. The inset shows the CMB temperature data at the Cold Spot.}
\end{center}
\end{figure*}

We show an image (without $\ell<10$ modes) and a radial profile (with and without $\ell<10$ modes) of the coldest spot in Jubilee and the corresponding radial $\Delta T_{\rm ISW}$ profile in degree units in the left panel of Figure 9. The imprint shows a cold spot in the center and a hot ring in the surrounding area, i.e. very similar to the shape of the Cold Spot and to the imprint of supervoids in general. We then compare the Jubilee coldest spot profiles to a single supervoid model estimate by \cite{Nadathur2014}, a multiple void model estimate by \cite{Mackenzie2017}, and to the statistical analysis of ISW contributions to coldest spots in simulated CMB+ISW maps \cite{Naidoo2017}. We report good consistency among all the results. 

However, a few additional forethoughts are needed before comparing this finding to the observational results. Models studied by \cite{Nadathur2014} and \cite{Mackenzie2017}, above all, fail to explain the hot ring in the shape of the Cold Spot profile and can only predict $\Delta T_{0} \approx-20-30 \mu K$ whereas the observed depression in the Cold Spot center is $\Delta T_{0} \approx -150 \mu K$. If the assumed void density profile is more strongly compensated, then the more general ISW model by \cite{FinelliEtal2014} can predict hot ring features around a central cold spot feature. Since the overall fluctuation in the $\Phi$ gravitational potential becomes less significant in those models \citep[see also][]{Naidoo2016}, consequently the magnitude of the central ISW imprint is reduced. Note that such modification in the modeling assumptions could bring the reconstructed ISW profiles of the Cold Spot closer to the profile of the coldest Jubilee spot.

For our main analysis, we conservatively choose the coldest spot profile defined in the $\ell\geq10$ Jubilee map. As discussed in Section 4.2, this way the profile is presumably free of temperature biases, caused by large-scale modes, at the expense of slightly reducing the magnitude of the signal. Note that this profile shape closely resembles the imprint of the largest Jubilee and BOSS supervoids.

\begin{figure*}
\begin{center}
\includegraphics[width=145mm]{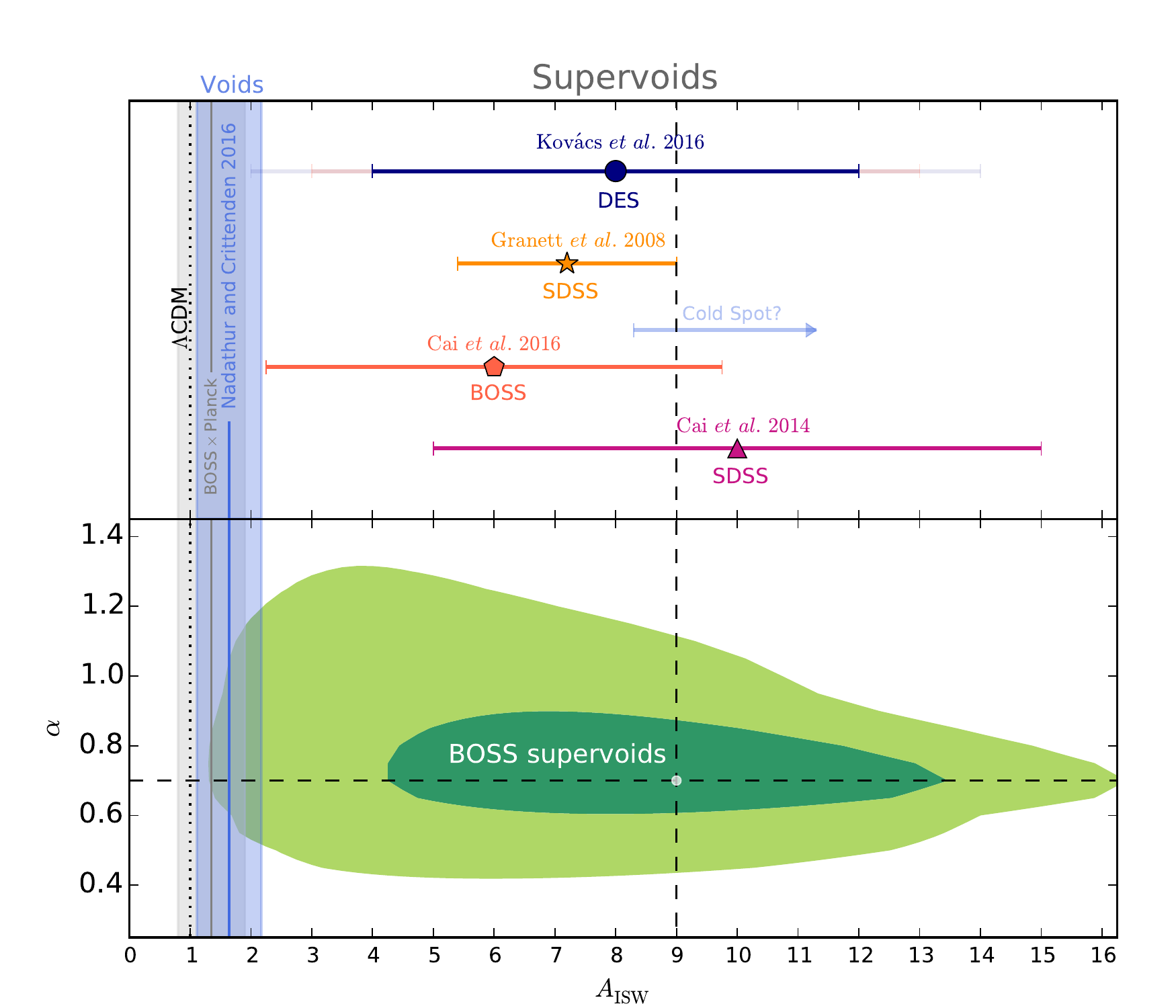}
\caption{Summary of our BOSS results and a comparison to other \emph{super}void-based $1\sigma$ estimates of the $A_{\rm ISW}$ amplitude. We assumed $\Delta T_{f}=-1.33~\mu K$, estimated by Nadathur et al. (2012), for the measurement by Granett et al. who used SDSS data. See Cai et al. (2014, 2016) for their simulation estimates of the $\Lambda$CDM signal. Light blue and light red error bars correspond to separate DES void and DES supercluster results, respectively, while the dark blue error bar is the combined DES constraint. The dark green and light green contours mark $1\sigma$ and $2\sigma$ confidence levels, respectively, around the best fit value with $A_{\rm ISW}\approx9$, $\alpha \approx0.7$. The dotted vertical black line marks the Jubilee $\Lambda$CDM expectation with $A_{\rm ISW}=1$. The gray band shows {\it Planck} $1\sigma$ constraints based on angular cross-correlations using BOSS data, while the light blue band corresponds to $1\sigma$ constraints by Nadathur $\&$ Crittenden (2016).}
\end{center}
\end{figure*}

The resulting $A_{\rm ISW}=1$ template coldest ISW spot profile, shown in the right panel of Figure 9, is of course in clear disagreement with the Cold Spot observations. If, however, the ISW amplitude is \emph{blindly} enhanced to $A_{\rm ISW}\approx9$ based on our BOSS findings, then the resulting central temperature depression closely matches that of the Cold Spot. We emphasize that the empirical relations based on our BOSS analyses play a key role in this comparison. 

The qualitative and quantitative agreement in the full extent of the profile, including the hot ring region, is remarkable. The agreement can be further improved by {\it a posteriori} increasing the angular size of the imprint by $20\%$ and by changing the amplitude to $A_{\rm ISW}\approx7.5$. It is important that the size of the particular coldest ISW spot can change from realization to realization thus such a difference is not unexpected. Besides, the BOSS \emph{super}void data is also consistent with slightly lower values of $A_{\rm ISW}$.

We note that features like the Cold Spot are also compatible with the statistical properties of coldest spots in random CMB maps thus chance correlation is a possible explanation \citep{Nadathur2014,Naidoo2017}. Proposing a chance alignment of a primordial cold spot and a $\Lambda$CDM ISW imprint of multiple voids, \cite{Naidoo2016} argued that subtracting the reconstructed ISW imprints from the observed Cold Spot profile reduces the extremeness of the Cold Spot below $\approx2\sigma$. 

Our logic here, however, was different. Instead of focusing on the Cold Spot as a CMB anomaly, we first analyzed BOSS \emph{super}voids and then finally estimated what the ISW\emph{-like} profile of the Eridanus \emph{super}void might look like if the observed $A_{\rm ISW}\approx9$ value is considered.

We conclude that, if the enhanced density-temperature correlation amplitude of \emph{super}voids is confirmed, the Cold Spot can be a further evidence for such an unexpected cosmological phenomenon or for a strange ISW-\emph{like} CMB contamination that is correlated with these large-scale density fluctuations.

\section{Conclusions}

We performed localized measurements of the ISW effect by analyzing the imprints of cosmic \emph{super}voids. With the Jubilee simulation, we critically revisited most of the aspects of the recent observational results by \cite{Cai2016} and \cite{NadathurCrittenden2016} who also analyzed BOSS DR12 void catalogues based on other void definitions. In our simulation analyses, we found that the largest $\approx20\%$ of the minimal voids show special ISW imprints with cold spots and surrounding hot rings. This way we defined a sample of \emph{super}voids prior to looking at the real-world data. 

We then used the simulated results as templates when looking for a signal in BOSS data. We detected, in disagreement with \cite{NadathurCrittenden2016} but in virtual agreement with \cite{Cai2016}, an excess ISW-\emph{like} void-temperature correlation signal with $A_{\rm ISW}\approx9$ at the $\approx2.5\sigma$ significance level. The tension with the Jubilee-based $\Lambda$CDM predictions appears to be $\gsim2\sigma$, as shown against other constraints in Figure 10 (based on various techniques used for theory and measurements thus comparisons are rather qualitative). An interesting feature identified in the BOSS data was the more compact shape of the imprint profile compared to Jubilee ISW reconstructions. We therefore needed an {\it a posteriori} re-scaling parameter $\alpha\approx0.7$ for a satisfactory matching between observed and simulated ISW profiles. This feature might point to the fragility of the signal, even if DES \emph{super}voids also showed a similar pattern \citep{Kovacs2016}.

Our results also provide a framework to re-think the case of the CMB Cold Spot. With a significant enhancement of the density-temperature correlation at large scales, the Cold Spot and the Eridanus \emph{super}void are plausibly related via this unexpected phenomenon.

Blaming systematic effects, one can think of residual contamination coming from unresolved extragalactic point sources that might still contaminate the ISW measurements and cosmological parameter estimation \citep[see e.g.][]{Millea2012}. Dust from galaxies at all redshifts contributes to the CMB temperature fluctuations, which, in turn, would result in a positive correlation between CMB temperatures and galaxy density \citep[see e.g.][]{ho}.

More speculatively, the excess signals can point to {\it new physics}. \cite{Nadathur2012} discussed that the freedom to vary the parameters of $\Lambda$CDM models, given other precise constraints, is not enough to overcome these discrepancies with observation. Non-Gaussianities in the primordial perturbations might result in excess ISW signals but {\it Planck} analyses have practically excluded this possibility \citep{Planck2015NonGauss}. 

Modified gravity theories with alternative growth rates, however, might provide some ground to discuss such excess signals. Possibly, the interesting discrepancy in the measured imprint profiles is related to spatial perturbations in dark energy that are expected to mainly alter large-scale physics and their unique ISW effect is their main hope to be uncovered \citep[see e.g.][]{WellerLewis2003,Bean2004,HuScranton2004,Mota2008,dePutter2010}.

In summary, our results suggest that it is interesting to consider a smaller catalogue of the largest voids instead of the largest catalogue of smaller voids for ISW studies. The origin of our observations remains unclear but certainly warrants further studies and alternative modeling techniques including modified gravity theories or implications to possible backreaction of large-scale structure \cite[see e.g.][]{Racz2016}, tests of the imprint of these superstructures in CMB lensing convergence maps, and connections to large-angle CMB anomalies \citep{Schwarz2015}.

\section*{Acknowledgments}
AK thanks Seshadri Nadathur for providing his void catalogues, and for useful discussions about pruning techniques and ISW analyses of mock voids. The author also thanks the Jubilee team for providing their LRG mock data and ISW map based on their N-body simulation that was performed on the Juropa supercomputer of the J\"ulich Supercomputing Centre (JSC). Comments by Istv\'an Szapudi, Juan Garc\'ia-Bellido, Ramon Miquel, Krishna Naidoo, Yan-Chuan Cai, John Peacock, and Nico Hamaus have also greatly improved the paper. Funding for this project was partially provided by the Spanish Ministerio de Econom\'ia y Competitividad (MINECO) under projects FPA2012-39684, and Centro de Excelencia Severo Ochoa SEV-2012-0234. The author has also been supported by a Juan de la Cierva fellowship from MINECO.

This work has made use of public data from the SDSS- III collaboration. Funding for SDSS-III has been provided by the Alfred P. Sloan Foundation, the Participating Institutions, the National Science Foundation, and the U.S. Department of Energy Office of Science. The SDSS-III website is http://www.sdss3.org/. 

\bibliographystyle{mn2e}
\bibliography{refs}
\end{document}